%% file: unsure2023_melba.tex
\documentclass[twoside,11pt]{article}

\usepackage[accepted,specialissue]{melba}

\usepackage{longtable}
\usepackage{multirow}

\usepackage{acronym}
\usepackage{booktabs}
\usepackage{graphicx}
\usepackage{array}
\usepackage{comment}

\acrodef{ASIR}[ASiR]{adaptive statistical iterative reconstruction}
\acrodef{DICOM}{Digital Imaging and Communications in Medicine}
\acrodef{CCC}{concordance correlation coefficient}
\acrodef{C-index}{Harrell's C-index}
\acrodef{CPH}{Cox proportional hazards}
\acrodef{CRLM}{colorectal liver metastases}
\acrodef{CT}{computed tomography}
\acrodef{IRB}{institutional review board}
\acrodef{LMM}{linear mixed model}
\acrodef{MSK}{Memorial Sloan Kettering Cancer Center (New York, NY)}
\acrodef{MDA}{MD Anderson Cancer Center (Houston, TX)}
\acrodef{MRMR}[mRMR]{minimum redundancy, maximum relevancy}
\acrodef{GLCM}{gray level co-occurrence matrix}
\acrodef{NGTDM}{neighboring gray-tone difference matrix}
\acrodef{GLDM}{gray level dependence matrix}
\acrodef{GLRLM}{gray level run length matrix}
\acrodef{GLSZM}{gray level size zone matrix}
\acrodef{IBSI}{Image Biomarker Standardisation Initiative}
\acrodef{NIFTI}[NIfTI]{Neuroimaging Informatics Technology Initiative}
\acrodef{RD}[R. D.]{R. Do}
\acrodef{ROI}{region of interest}
\acrodef{SimpleITK}[Simple ITK]{Simple ITK}
\acrodef{TCIA}{the Cancer Imaging Archive}
\acrodef{XNAT}{Extensible Neuroimaging Archive Toolkit}
\acrodef{HU}{Hounsfield units}

\newacroplural{ROI}{regions of interest}

\acused{RD}
\acused{SimpleITK}

\newcommand{\numpat}{81}
\newcommand\msknumpat{44}
\newcommand\mdanumpat{37}

\newcommand\tabref[1]{Table~\ref{#1}}
\newcommand\figref[1]{Figure~\ref{#1}}

\melbaid{2024:032}  %
\doi{https://doi.org/10.59275/j.melba.2024-24gc}
\melbaauthors{Peoples et al}  %
\volume{2}
\firstpageno{2326}  %
\melbayear{2024}  %
\datesubmitted{01/2024}  %
\datepublished{01/2025}  %

\melbaspecialissue{Uncertainty for Safe Utilization of Machine Learning in Medical Imaging \\ (UNSURE) 2023}
\melbaspecialissueeditors{Christian Baumgartner, Adrian Dalca, Raghav Mehta, Chen Qin, Carole Sudre, \\ William (Sandy) Wells}

\ShortHeadings{Finding Reproducible and Prognostic Radiomic Features}{Peoples et al.}

\title{Finding Reproducible and Prognostic Radiomic Features in Variable Slice Thickness Contrast Enhanced CT of Colorectal Liver Metastases}

\author{%
\name Jacob J. Peoples\orcid{0000-0003-0191-7446}
    \email jacob.peoples@queensu.ca \\
    \addr School of Computing, Queen's University, Kingston, ON, Canada   
\AND
\name Mohammad Hamghalam\orcid{0000-0003-2543-0712} \\
    \addr School of Computing, Queen's University, Kingston, ON, Canada \\
    \addr Department of Electrical Engineering, Qazvin Branch, Islamic Azad University, Qazvin, Iran
    \AND
\name Imani James \\
    \addr Department of Radiology, Memorial Sloan Kettering Cancer Center, New York, NY, USA
    \AND
\name Maida Wasim \\
    \addr Department of Radiology, Memorial Sloan Kettering Cancer Center, New York, NY, USA
    \AND
\name Natalie Gangai\orcid{0000-0003-0603-7165} \\
    \addr Department of Radiology, Memorial Sloan Kettering Cancer Center, New York, NY, USA
    \AND
\name Hyunseon Christine Kang\orcid{0000-0001-5313-3526} \\ 
    \addr Department of Abdominal Imaging, The University of Texas MD Anderson Cancer Center, Houston, TX, USA
    \AND
\name X. John Rong\orcid{0000-0002-1169-5042} \\ 
    \addr Department of Imaging Physics, The University of Texas MD Anderson Cancer Center, Houston, TX, USA 
    \AND
\name Yun Shin Chun\orcid{0000-0003-1384-8927} \\
    \addr Department of Surgical Oncology, The University of Texas MD Anderson Cancer Center, Houston, TX, USA
    \AND 
\name Richard K. G. Do\orcid{0000-0002-6554-0310} \\ 
    \addr Department of Radiology, Memorial Sloan Kettering Cancer Center, New York, NY, USA
    \AND
\name Amber L. Simpson\orcid{0000-0002-4387-8417} \\
    \addr School of Computing, Queen's University, Kingston, ON, Canada \\
    \addr Department of Biomedical and Molecular Sciences, Queen's University, Kingston, ON, Canada
}

\begin{document}

\maketitle

\begin{abstract}%
Establishing the reproducibility of radiomic signatures is a critical step in the path to clinical adoption of quantitative imaging biomarkers; however, radiomic signatures must also be meaningfully related to an outcome of clinical importance to be of value for personalized medicine. In this study, we analyze both the reproducibility and prognostic value of radiomic features extracted from the liver parenchyma and largest liver metastases in contrast enhanced CT scans of patients with colorectal liver metastases (CRLM). A prospective cohort of 81 patients from two major US cancer centers was used to establish the reproducibility of radiomic features extracted from images reconstructed with different slice thicknesses. A publicly available, single-center cohort of 197 preoperative scans from patients who underwent hepatic resection for treatment of CRLM was used to evaluate the prognostic value of features and models to predict overall survival. A standard set of 93 features was extracted from all images, with a set of eight different extractor settings. The feature extraction settings producing the most reproducible, as well as the most prognostically discriminative feature values were highly dependent on both the region of interest and the specific feature in question. While the best overall predictive model was produced using features extracted with a particular setting, without accounting for reproducibility, (C-index = 0.630 (0.603--0.649)) an equivalent-performing model (C-index = 0.629 (0.605--0.645)) was produced by pooling features from all extraction settings, and thresholding features with low reproducibility ($\mathrm{CCC} \geq 0.85$), prior to feature selection. Our findings support a data-driven approach to feature extraction and selection, preferring the inclusion of many features, and narrowing feature selection based on reproducibility when relevant data is available.
\end{abstract}


\section{Introduction}

Radiomic analysis as a field is predicated on the idea that radiological imaging contains meaningful biological information contained in the patterns of intensity values within regions of interest, which could contribute to a better understanding of patient health~\citep{gillies2016radiomics}.
The typical approach in radiomic studies is to extract a large number of pre-defined quantitative imaging features from a region of interest, and then use machine learning methods to reduce the dimensionality of the feature set and build models of a biological correlate or a medical outcome of interest~\citep{horvat2022primer}.
The promise of radiomics to develop quantitative imaging biomarkers is of broad interest because it poses a non-invasive means to characterize patient disease using routine clinical imaging.
However, to be clinically deployed, radiomic models must be widely validated, and their robustness to variable imaging settings well-established.

In the present study, we are focused on contrast-enhanced abdominal \ac{CT} of the liver for patients with \ac{CRLM}.
These patients have an overall poor prognosis, which could potentially be improved by prognostic radiomic signatures that could better target patients for surgery or chemotherapy.
In patients with liver metastases, radiomic models derived from contrast-enhanced \ac{CT} have shown substantial prognostic capability in both survival modeling and prediction of chemotherapy response~\citep{fiz2020radiomics}.
Furthermore, although most studies focus on radiomic features extracted from the metastases themselves, studies of \ac{CRLM} have shown that features from the liver parenchyma also contain important information when predicting hepatic disease-free survival or overall survival after hepatic resection~\citep{simpson2017computed}, or progression-free survival after radiotherapy~\citep{hu2022radiomics}.
With these applications in mind, in this study we are seeking to gain a better understanding of the robustness of radiomic models in contrast-enhanced \ac{CT} of \ac{CRLM}.

One way to study the robustness of radiomic models is to study the reproducibility of the model inputs---that is, to understand how consistent the radiomic features are under real-world variations that occur in image acquisition and reconstruction.
Treating the radiomic features as measurements drawn from radiological images, the reproducibility of the features can be studied from a metrological perspective~\citep{raunig2014quantitative}, using statistical measures such as the \ac{CCC}.
Understanding the reproducibility of radiomic features, however, is difficult given the many factors affecting the results.
Restricting the discussion to \ac{CT}, scans are acquired from different medical centers, using different scanners, with different imaging acquisition parameters and protocols, after which the images are reconstructed with different algorithms, different resolution parameters, and different kernels, all of which have been shown to affect feature reproducibility~\citep{zhao2021understanding}. 
Once the images are in hand, feature extraction itself is not without concerns regarding reproducibility; while standardization efforts such as the \ac{IBSI}~\citep{zwanenburg2020image} have been instrumental in creating a common nomenclature and a standard set of well-defined features, they do not provide a final answer in how to set the many configurable parameters which can affect the final feature values computed by common software such as \texttt{pyradiomics}~\citep{vanGriethuysen2017computational}.

In this study, we are considering feature reproducibility particularly when the slice thickness used to reconstruct the \ac{CT} scans is varied.
When comparing features extracted from \ac{CT} images with thinner or thicker slices, studies have found that features from thinner images are more reproducible across variations in segmentation~\citep{hu2022effects}, or across repeat imaging~\citep{zhao2016reproducibility},
and may also produce more accurate models~\citep{he2016effects,li2018ct,xu2022effect}.
In direct comparisons, many features had poor reproducibility when comparing features from images with different slice thicknesses in a variety of phantom studies~\citep{zhao2014exploring,ger2018comprehensive,berenguer2018radiomics,kim2019effect,varghese2019reliability,ligero2021minimizing,ibrahim2022maaspenn}.
Prospective studies producing multiple reconstructions for each patient have reproduced this result on patient images for
lung cancer~\citep{lu2016assessing,park2019deep,erdal2020quantitative,yang2021evaluation,emaminejad2021reproducibility},
and liver metastases~\citep{meyer2019reproducibility}.
Poor feature reproducibility with respect to slice thickness is concerning because many real-world retrospective or multi-site data sets include images with a range of different slice thicknesses due to variations in local protocols~\citep{ger2018comprehensive}.
Although image interpolation to a common, isotropic voxel size is considered a best practice for preprocessing during feature extraction~\citep{zwanenburg2020image} in order to ensure the image features are comparable between images with different voxel sizes, the optimal choice of resolution and resampling algorithm is undecided.
Furthermore, resampling to a common voxel size appears, on its own, to be insufficient to overcome the inconsistency of feature values due to slice thickness variation, except in a small subset of features~\citep{shafiq-ul-hassan2017intrinsic,shafiq-ul-hassan2018voxel}.

To further complicate matters, it has been shown that reproducibility is not necessarily consistent across cancer types, even for a single modality such as \ac{CT}~\citep{vanTimmeren2016test-retest}.
However, in a systematic review of radiomics reproducibility studies, \cite{traverso2018repeatability} found that the literature was limited to a small number of cancer types, with the greatest number of studies addressing lung cancers, amongst which \ac{CT} was the most common imaging modality.
Across these studies there was not a clear consensus on the most reproducible features, although in \ac{CT} they found agreement that first-order features tended to be more reproducible than higher-order texture features.
Ultimately, it seems that the reproducibility of features is not easily generalized across different anatomies or cancers.
Therefore, reproducibility studies for the \ac{ROI} and disease under consideration is an important part of the validation any radiomics-based imaging biomarker.

In this study we present an analysis of the relationship between the reproducibility and prognostic value of radiomic features drawn from contrast enhanced \ac{CT} of \ac{CRLM}.
We present a reproducibility analysis of radiomic features on a cohort of 81 prospectively enrolled patients from two major US cancer centers, who underwent contrast enhanced abdominal imaging with a controlled and systematically varied protocol.
Our analysis is primarily focused on the effects of slice thickness chosen at reconstruction time.
Features were extracted from the largest liver metastasis and the liver parenchyma from each subject using a variety of different configurations, varying the level of resampling, and the method of aggregation used in computing the higher-order texture features.
To investigate the relationship between reproducibility and prognostic value, we conducted an in depth univariate and multivariable survival modeling analysis on an independent, publicly available data set of 197 preoperative contrast enhanced \ac{CT} scans of patients who underwent hepatic resection to treat \ac{CRLM}~\citep{simpson2023preoperative,simpson2024preoperative}.

This paper is a significant expansion of our previous work~\citep{peoples2023examining}, with a greater focus on the relationship between the reproducibility and prognostic value of the radiomic features under consideration.
In this paper, we are less concerned with finding the feature extraction settings that produce the most reproducible features.
Instead, we focus on the integration of reproducibility information into the development of prognostic radiomic signatures.
We present a joint, univariate analysis of both the reproducibility and prognostic discriminative ability of the features, taking a multi-objective optimization point of view.
The multivariable analysis was revised to use a standard feature selection algorithm, and to conduct many iterations of the cross-validation to ensure our results were stable.
Additional details are provided throughout the paper, giving greater context on the methodology, more visualisation and discussion of the results, and more interpretation of how our results fit into the greater context of the literature on the reproducibility of radiomics.
Code for all analysis is available at \url{github.com/jpeoples/melba2024}.

\section{Methods}

\subsection{CT Imaging and Segmentation}

Contrast enhanced, portal venous phase \ac{CT} scans were prospectively collected from a total of \numpat{} patients with \ac{CRLM} from two institutions, \ac{MSK} (n=\msknumpat{}) and \ac{MDA} (n=\mdanumpat{}), with \acl{IRB} approval and informed consent.\footnote{%
Note that all scans used to generate the results in this paper are consistent with the standard clinical acquisitions at each site, that would have been acquired for these patients regardless of their involvement in the study. Informed consent was required because the full study protocol includes a second scan, taken just before or after the usual scan (within $\pm 15$~s), to capture the test-retest repeatability of radiomic features in the presence of contrast. These results are not included in the present paper because the add-on scans were not processed at the time of preparation.  
}
Every scan was collected on a multi-detector CT scanner (Discovery CT750 HD; GE Healthcare, Madison, WI, USA) with 64 detector rows, and 0.625~mm detector width, for a total collimation width of 40~mm.
The images were collected with a tube voltage of 120 kVp, and automated tube current modulation using GE Smart mA with a noise index of 14 (\ac{MSK}) or 11 (\ac{MDA}). The tube current range varied between centers, with \ac{MSK} using range 220-380 mA, while \ac{MDA} used range 275-650 mA.
The gantry rotation time was 0.7~s with pitch factor 0.984 for \ac{MSK}, and 0.5 s with pitch factor 0.516 for \ac{MDA}.
All images were reconstructed with the standard soft tissue convolution filter.

To study the reproducibility of radiomic features with respect to image reconstruction, each patient scan was retrospectively reconstructed with different slice thicknesses and levels of \ac{ASIR} after image acquisition. 
In particular, every combination of three different slice thicknesses---2.5~mm, 3.75~mm, 5~mm---and seven different levels of \ac{ASIR}---from 0\% (equivalent to filtered back-projection), to 60\% in increments of 10\%---were generated, giving a total of $3\times 7 = 21$ reconstructions for every scan.
In all cases, the slice thickness and slice interval were equal, such that the reconstructed slices were spatially contiguous, and non-overlapping.\footnote{%
Because the CT scanners in this study (GE Discovery CT750) have an equal number of detectors and channels, the thinnest possible slices (corresponding to a single detector row width of 0.625~mm) are always acquired, and the slice thickness is applied only at reconstruction time, grouping data from multiple detector rows to produce thicker slices.
In our case, 2.5~mm, 3.75~mm, and 5~mm slice thicknesses correspond to grouping 4, 6, or 8 individual 0.625~mm rows.
}
Images were stored and transferred in \ac{DICOM} format after deidentification from both \ac{MSK} and \ac{MDA}. All \ac{CT} data was converted into \ac{NIFTI} format for further processing.

\subsection{Segmentation}
A single reconstruction (slice thickness 5~mm and \ac{ASIR} 20\%) was chosen as the reference reconstruction for each patient, for manual segmentation verification and correction by an experienced radiologist (\ac{RD}).
The choice of 5~mm and 20\% \ac{ASIR} for the reference was made to more closely match the standard-of-care clinical imaging protocol at the radiologist's home institution (\ac{MSK}).
The segmentation of these reference scans was completed in two phases: first, an automated segmentation was generated, after which the radiologist verified and corrected each mask in 3D Slicer~\citep{kikinis20133d}.
The initial segmentations were generated using an nnU-net model~\citep{isensee2021nnu-net} trained on a public database of 197 \ac{CT} scans from patients with \ac{CRLM}~\citep{simpson2023preoperative,simpson2024preoperative}, available from \ac{TCIA}~\citep{clark2013cancer}.
Details on the development of this model can be found in several previous publications \citep{hamghalam2021modality,mojtahedi2022towards,hamghalam2023attention-based}.
After the radiologist corrected the segmentations for the reference reconstruction, segmentations for other slice thicknesses were generated by resampling the reference segmentation using nearest-neighbor interpolation with the \ac{SimpleITK} software library.
No changes were made to the segmentations for reconstructions with different \ac{ASIR}, given that the tissue being imaged does not change in a given slice between \ac{ASIR} levels.

\subsection{Radiomic Feature Extraction}

Features were extracted from every image using \texttt{pyradiomics}~\citep{vanGriethuysen2017computational}, which is one of several open-source packages implementing a large set of \ac{IBSI}-compliant features.
Two 3D \acp{ROI} were used: the largest tumor within the liver, and the liver parenchyma (with all tumors and vessels excluded).
The \texttt{pyradiomics} library has seven default classes of features: shape, first order, \ac{GLCM}, \ac{GLDM}, \ac{GLSZM}, \ac{GLRLM}, \ac{NGTDM}.
In this study we used all feature classes, with the exception of shape features, which were excluded because there was only one reference segmentation per patient, and therefore any variations across reconstructions would only reflect the effects of interpolation.
From the remaining six classes, all default features were included.
Although \texttt{pyradiomics} supports the extraction of features from a variety of derived images, in addition to the original intensity image, we did not include any analysis of these derived image features.
The total number of features used in this study is broken down by feature class in \tabref{tab:featurecounts}, and a complete list is provided in \tabref{tab:feature_list} in the appendix.

\begin{table}[t]
\caption{\label{tab:featurecounts} Feature counts by class.}
\centering
\input{feature_counts_by_class.txt}

\end{table}

\subsubsection{Terminology}

Before continuing, we will establish some key terminology used throughout the remainder of this paper.
We will frequently refer to two classes of radiomic features: first-order features, and texture features.
For our purposes, first-order features are those that are computed strictly from the intensity histogram for all voxels in the \ac{ROI}, and corresponds to the first order feature class in \texttt{pyradiomics}.
Texture features, on the other hand, will be used strictly to refer to the remaining classes of higher-order features---\ac{GLCM}, \ac{GLDM}, \ac{GLSZM}, \ac{GLRLM}, and \ac{NGTDM}---which share the common factor of accounting for relationships between the intensities of neighboring voxels.
Note that this usage mirrors the terminology used by \ac{IBSI}~\citep{zwanenburg2020image}.

\subsubsection{Preprocessing}
A typical radiomic feature extraction pipeline includes a number of image preprocessing steps prior to feature computation. 
In our feature extraction, there were three preprocessing steps that every image underwent: image resampling, mask resegmentation, and intensity discretization.

Image resampling refers to the process of altering the input image resolution using an interpolation process.
Given that texture features take account of relationships between neighboring voxels, resampling is important, because otherwise the neighbors being compared would not be an equal physical distance apart across images with different resolution, changing the meaning of the feature.
To resolve this issue, the \ac{IBSI} reference manual~\citep{zwanenburg2016image} recommends resampling images to a common resolution.
Furthermore, the \ac{IBSI} manual recommends a common \emph{isotropic} resampling, in order to ensure the rotational invariance of 3D texture features, which account for relationships between neighboring voxels in all 3D directions.
Because radiomic features depend on both the underlying \ac{CT} image, and a segmentation of the \ac{ROI}, when applying resampling, both the image and segmentation mask must be interpolated.

Mask resegmentation refers to the removal of voxels outside of a preconfigured range when computing the first-order and texture features.
Although \cite{zwanenburg2016image} do not give specific recommendations about the optimal settings for resegmentation, it is commonly applied to ensure that outlier intensity values (due to small errors in the segmentation mask, or due to artifacts) do not skew the resulting feature distributions.

Intensity discretization (or quantization) is a binning process used to reduce the number of unique intensity values in the image, which is used in the computation of texture features, as well as some first-order features which require a probability density based on the intensity histogram.
Discretization is known to have a substantial effect on feature values~\citep{shafiq-ul-hassan2017intrinsic}, although the effect on feature reproducibility may be limited~\citep{larue2017influence}.

In this study, all features were computed using a discretization level of 24 bins.
The masks were resegmented using a window of $[-50, 350]$ \ac{HU}, in order to exclude metal artifacts (stents, etc), as well as rare and implausible outlier intensities.
All \ac{CT} images were interpolated using  B-splines, while the segmentation masks were interpolated using nearest-neighbor interpolation, which are the default algorithms in \texttt{pyradiomics}.
We tested several resampling resolutions, which are described in greater detail below.

\subsubsection{Texture Feature Aggregation}

The \ac{IBSI} reference manual~\citep{zwanenburg2016image} breaks feature aggregation into three categories---2D, 2.5D, and 3D.
The exact details of how the aggregation works varies across the classes of texture features, but the key factors are as follows.

\paragraph{3D vs 2D or 2.5D:} All texture features consider the relationships between neighboring voxels. The first key factor in the aggregation methods is that while 3D aggregation includes neighbors from any direction in three dimensions, 2D and 2.5D aggregation only considers neighboring voxels within the same axial plane.

\paragraph{2.5D vs 2D:} The second key factor differentiates 2D and 2.5D methods. 
Each class of texture feature is defined by an underlying matrix representing some aspect of the relationships between neighboring voxels.
Individual features are then defined in terms of equations operating on this matrix.
For 2D aggregation, the matrix is computed independently for each slice, and the resulting features are merged across slices by averaging.
In the case of 2.5D aggregation, a single matrix is computed, which includes relationships from all slices, which is then used to compute the features in the usual way.
This differs from 3D aggregation in that for 2.5D aggregation, only neighbors which share the same axial plane are considered, while in 3D aggregation, all neighbours in all three dimensions are included.

\paragraph{Directional vs non-directional:} Another factor is whether the matrix underlying the feature class is defined directionally or not.
The \ac{GLCM} and \ac{GLRLM} are both defined per direction.
Taking symmetry into account, this means that there are four or thirteen unique \ac{GLCM}/\ac{GLRLM} in 2D or 3D respectively.
Therefore, the matrices for each direction can be used to derive directional features, which can then be averaged, or the matrices can be merged via addition and used to compute single features.
In the present work, we rely on the former approach for both 3D and 2.5D aggregated features, which corresponds to the \ac{IBSI}-defined classes ITBB and JJUI, respectively.

The matrices underlying the remaining classes---\ac{GLDM}, \ac{NGTDM}, \ac{GLSZM}---consider neighbors in all included directions in a single matrix.
Therefore, no directional merging takes place.
For these, 2D aggregation would compute a matrix per slice, which would then be used to derive per-slice features, which could be averaged; 2.5D aggregation would merge the 2D matrices across all slices, before computing features; 3D aggregation would compute a single matrix for the entire \ac{ROI}, because neighbors in all directions are counted.
In the present work we are using both 2.5D and 3D aggregation, which correspond to the \ac{IBSI}-defined classes 62GR and KOBO, respectively.

A breakdown of the texture feature classes and the aggregation methods we are using in the present study is given in \tabref{tab:aggregation}.

\begin{table}
\caption{\label{tab:aggregation} Aggregation methods by feature class.}
\centering
\begin{tabular}{l l l l l}
\toprule
              &   \multicolumn{2}{c}{Aggregation} \\   
              \cmidrule{2-3}
Type & 2.5D & 3D & Feature Classes\\ \midrule

Directional& JJUI & ITBB & \ac{GLCM}, \ac{GLRLM} \\
Non-directional & 62GR & KOBO & \ac{NGTDM}, \ac{GLDM}, \ac{GLSZM}\\
 \bottomrule
\end{tabular}
\end{table}

\subsubsection{Resampling and Aggregation Variants}
\begin{figure}
    \centering
    \includegraphics[width=0.5\linewidth]{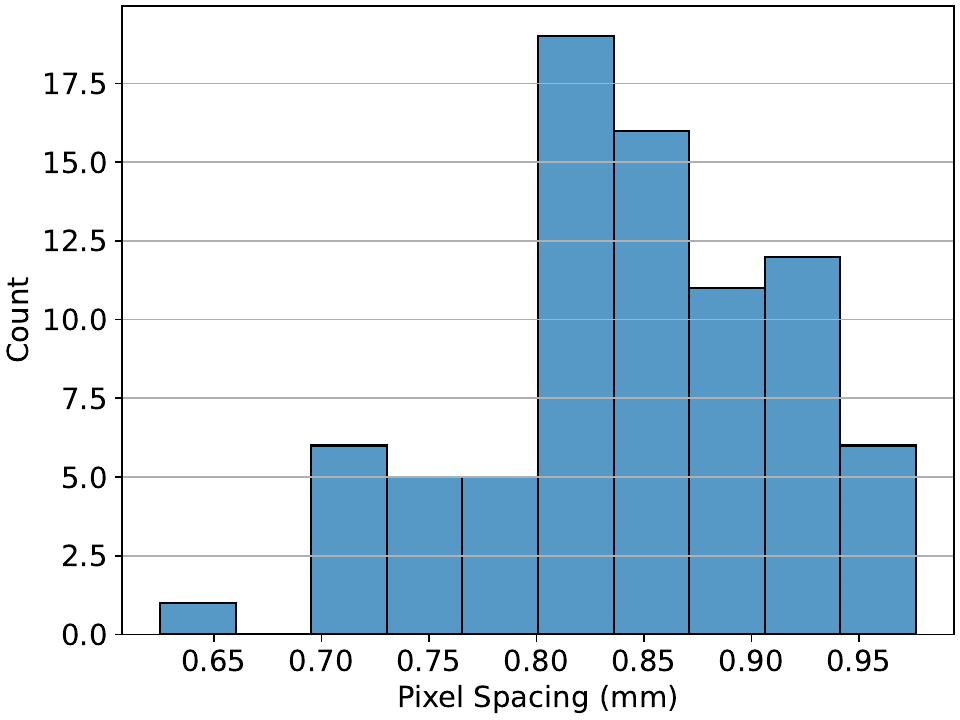}
    \caption{\label{fig:pixelsize} A histogram of the in-plane pixel spacing for images in our dataset. Pixel spacing was consistent across all reconstructions, so only counts for the reference reconstructions (5~mm slice thickness and 20\% ASiR) are shown.}
\end{figure}

Because our images are substantially anisotropic, with a larger axial slice thickness than in-plane pixel size, we wanted to investigate the optimality of resampling to an isotropic voxel size per the \ac{IBSI} recommendations~\citep{zwanenburg2016image}.
Accordingly, we investigated three different levels of resampling: $1\times 1 \times 1$~mm, $0.85\times 0.85 \times 0.85$~mm, and $0.85 \times 0.85 \times 2.5$~mm, where 1~mm was chosen as a typical value from the radiomics literature, 0.85~mm was chosen as the median in-plane pixel spacing in the data set, and 2.5~mm was chosen as the $10^{\mathrm{th}}$-percentile z-axis spacing, in analogy to the nnU-net resampling method for anisotropic imaging data sets~\citep{isensee2021nnu-net}.
The original distribution of in-plane pixel sizes in the data set prior to resampling is shown in \figref{fig:pixelsize}.

We also chose to extract features with both 3D and 2.5D feature aggregation at all resampling levels, to investigate whether 2.5D aggregation might be more suitable in anisotropic imaging.
Because 2.5D aggregation never includes comparisons of voxels in neighboring axial planes, maintaining a common z-axis resolution is also not necessary for features to remain comparable.
Therefore, in addition to the three previously described resampling levels, we also tried resampling the images to 0.85~mm or 1~mm in-plane resolutions, while preserving the original z-axis resolution, and extracting only 2.5D aggregated features.

All the aforementioned configurations led to a total of eight distinct feature extraction settings, which are summarized in \tabref{tab:feature_extraction}.
The table also introduces a naming scheme for these feature extraction settings, in order to enable the comparison of results across methods.
In the naming scheme, the voxel resolution is indicated by the letters ``L'' (large, i.e. 1~mm), ``S'' (small, i.e., 0.85~mm), and ``A'' (anisotropic, i.e. $0.85\times 0.85\times 2.5$~mm).
The feature aggregation is indicated by a number ``2'' or ``3'', for 2.5D or 3D aggregation, respectively.
Finally, a lower-case ``i'' is appended to the name to indicate that the resampling was restricted to in-plane only, preserving the z-axis resolution.

\begin{table}
\caption{\label{tab:feature_extraction}  Feature extraction methods.}
\centering
\begin{tabular}{l l l l}
\toprule
     & \multicolumn{2}{c}{Resampling (mm)} & \\
     \cmidrule{2-3}
Name & In-plane & z-axis & Aggregation \\
\midrule
L2i & 1 & None & 2.5D \\
L2  & 1 & 1 & 2.5D \\
L3  & 1 & 1 & 3D \\
S2i & 0.85 & None & 2.5D \\
S2  & 0.85 & 0.85 & 2.5D \\
S3  & 0.85 & 0.85 & 3D \\
A2  & 0.85 & 2.5 & 2.5D \\
A3  & 0.85 & 2.5 & 3D \\
\bottomrule
\end{tabular}
\end{table}

\subsection{Reproducibility Analysis}

The \ac{CCC}~\citep{lin1989concordance} was used to measure reproducibility of radiomic features.
The first phase of the analysis was restricted to the reference \ac{ASIR} level of 20\%. 
The standard pairwise \ac{CCC} was computed for every feature across all pairs of slice thicknesses (2.5~mm vs. 3.75~mm, 2.5~mm vs. 5~mm, 3.75~mm vs 5~mm).
The paired Wilcoxon sign-rank test~\citep{wilcoxon1945individual} was used to test the statistical significance of the change in \ac{CCC} between slice thicknesses.

The second phase of the analysis used a \ac{LMM} for each feature in order to compute a generalized \ac{CCC} using the data from all three slice thicknesses~\citep{carrasco2003estimating}.
In this model, the reconstructions with different \ac{ASIR} levels were also included, and controlled for as a fixed-effect when computing the \ac{CCC}.
In brief, using this approach, an \ac{LMM} is computed for each feature, which is used to estimate the variance due to subject, $\sigma^2_s$, the variance due to both slice thickness, $\sigma^2_t$ and \ac{ASIR}, $\sigma^2_a$, along with an error term, $\sigma^2_e$.
Following \cite{carrasco2003estimating}, the generalized \ac{CCC} is then
\begin{equation}
\mathrm{CCC} = \frac{\sigma^2_s}{\sigma^2_s + \sigma^2_t + \sigma^2_e}.
\end{equation}
For each feature, we used this method to compute a spectrum of \acp{CCC}, across all feature extraction settings, and \acp{ROI}.

\subsection{Survival Analysis on Independent Data Set}
\begin{figure}
    \centering
    \includegraphics[width=0.5\linewidth]{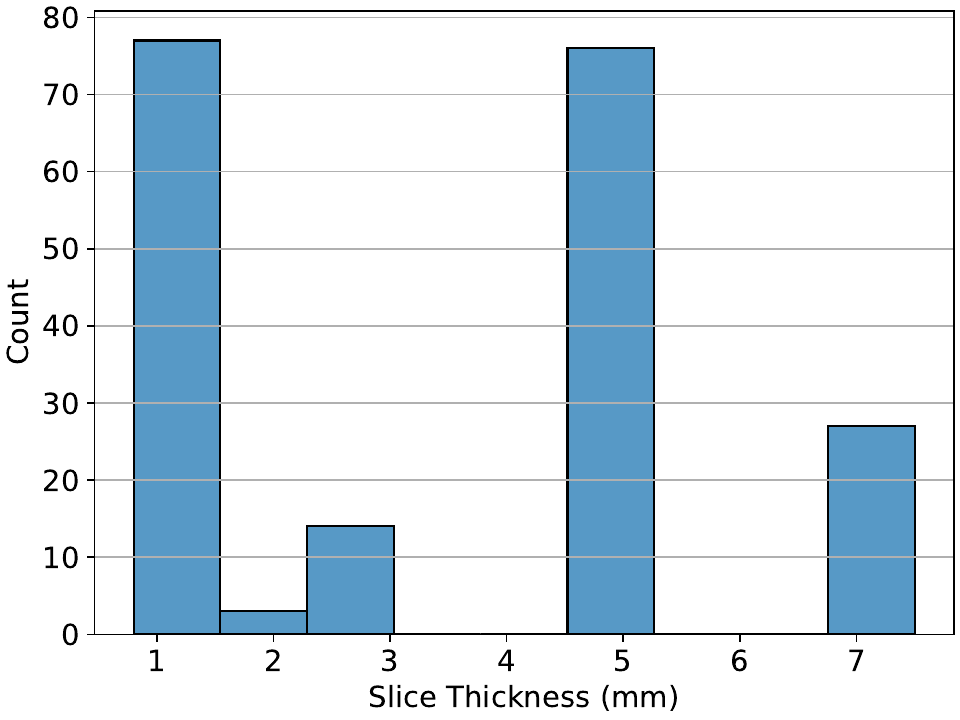}
    \caption{\label{fig:slicethickness} A histogram of the slice thickness for images in the survival data set. Slice thicknesses fell in the range $[0.8,7.5]$~mm.}
\end{figure}

The reproducibility of a feature is an independent consideration from its value as a predictor in a given context.
Ultimately, in radiomics the goal is to model a given outcome, and therefore, the predictive or prognostic values of features can not be sacrificed in order to create reproducible radiomic signature.
To address this concern we performed survival analysis on the aforementioned public data set of 197 \ac{CRLM} patients~\citep{simpson2023preoperative, simpson2024preoperative}.
In this data set, the reproducibility of features across slice thickness is an important consideration, because thickness varied widely across scans (range $[0.8,7.5]$~mm, see \figref{fig:slicethickness}). %
All scans in this data set were acquired prior to a hepatic resection to treat \ac{CRLM}, and the repository includes right-censored data on overall survival time post-operation.
For each of these pre-treatment scans, we extracted features using all eight of the different feature extractor settings previously described.
In each case, the settings used to configure \texttt{pyradiomics} were identical to those used on the reproducibility dataset.

We completed the survival analysis in two phases, the first looking at univariate relationships of the radiomic features with overall survival, and the second considering multivariable models.
In the first phase we computed the discriminative ability of each individual feature with respect to overall survival, as measured by Harrell's C-index~\citep{harrell1996multivariable}.
Features with negative discriminative ability ($\mathrm{C\mbox{-}index} < 0.5$) were negated, ensuring all C-index values fell in the range $[0.5,1]$.

In the second phase of the analysis, we did a repeated 10-fold cross-validation of a multivariable \ac{CPH} model building procedure.
Features from both \acp{ROI} were combined into one larger set for feature selection.
We first restricted the candidate features to those with $\ac{CCC} \geq \mathrm{CCC}_t$ in the reproducibility analysis, where $\mathrm{CCC}_t$ was a predetermined threshold. 
The model building process in each iteration then consisted of a feature selection, followed by \ac{CPH} modeling.
Features were selected by computing the univariate C-index for each feature, as described above, and removing any with $\mathrm{C\mbox{-}index} < 0.55$.
After this, univariate \ac{CPH} models were constructed for each remaining feature, and removed whenever the feature was not significant with $p < 0.1$.\footnote{%
We used a weak threshold for significance, because this is simply a filtering step.
}
Finally, the remaining features were reduced down to a predetermined number of features using the \ac{MRMR} feature selection algorithm~\citep{ding2005minimum}.
The 10-fold cross validation was repeated 100 times in order to get a stable estimate of the performance of the resulting multivariable models.
In this case, we did not compute a final model, because our goal was only to compare the performance across the models.

The entire multivariable modeling process was conducted for every feature extraction setting individually, as well as for the case of all features from all extraction settings combined into one large set.
For each feature set, the process was repeated for each \ac{CCC} threshold in the set $\mathrm{CCC}_t \in \{0, 0.8, 0.85, 0.9, 0.95\}$.
Finally, for each feature set, and each \ac{CCC} threshold, the process was repeated for every feature count in the set $\{1, 2, 4, 8, 16, 32, 64\}$.
This resulted in a total of
$
9 \text{ feature sets} \times 5 \text{ \ac{CCC} thresholds} \times 7 \text{ feature counts} = 315
$
different multivariable model cross-validation experiments.

\subsection{Statistical Analysis}

\subsubsection{Hierarchical Clustering}
To find patterns in the reproducibility and univariate predictive value of features, we used a hierarchical clustering and dendrogram visualization~\citep{mullner2011modern, bar-joseph2001optimal}.
For the reproducibility analysis, the rows of the matrix corresponded to the features, while the columns corresponded to each combination of \ac{ROI} and feature extractor, and the values were the \ac{CCC}.
By including both ROIs in the analysis, we were able to see clusters of patterns across and between \acp{ROI}.
For the univariate survival analysis, a similar clustering approach was used, although the two \acp{ROI} were split, because the same feature in either \ac{ROI} may have a different biological relevance to survival.
Therefore, in the univariate survival case, the rows corresponded to each feature, while the columns corresponded to the feature extractors, and the analysis was repeated for both \acp{ROI}.
In all cases, the hierarchical clustering was done for both the rows and columns, using Ward linkage~\citep{ward1963hierarchical}.

\subsubsection{Pareto Efficiency and Pareto Front}
In multi-objective optimization, a potential solution $A$ \emph{Pareto dominates} solution $B$ if for every objective, $A$ is better than $B$, or equally good.
A solution $A$ is \emph{Pareto efficient} if there exists no other solution that Pareto dominates it.
The \emph{Pareto front} refers to the set of all Pareto efficient solutions.
In other words, the Pareto front is the set of all solutions for which no improvement on any objective is possible, without a deterioriation of some other objective.
In order to better understand the relationship between the reproducibility and prognostic value of the features, we considered the set of features that are Pareto efficient over both the \ac{CCC} and univariate C-index when features were grouped across all extractors.
To understand the relationships between extractors, we also computed a Pareto front across feature extractors for each individual feature.

\section{Results}
\begin{figure}[!hb]
    \centering
    \includegraphics[width=\linewidth]{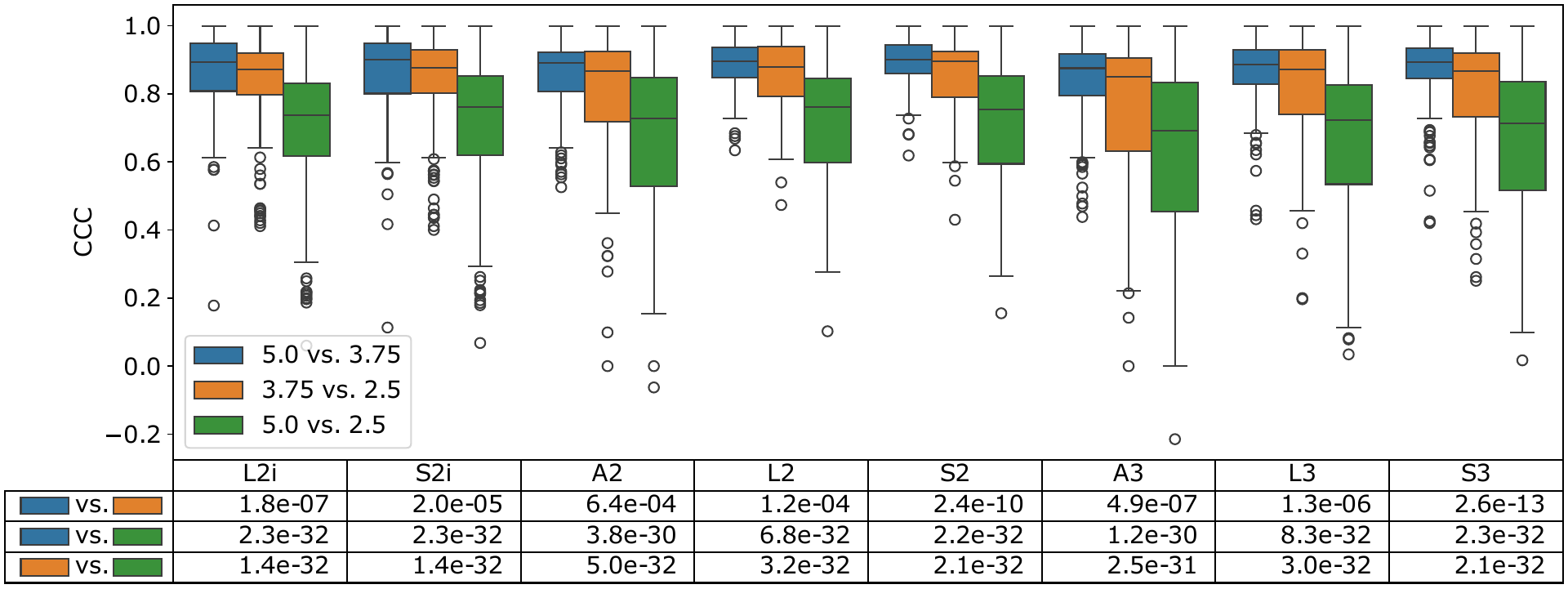}
    \caption{\label{fig:ccc_pairwise} Box plots of the features compared pairwise across slice thicknesses, broken down by feature extraction setting. The statistical significance of the change in \ac{CCC} value between different pairs of slice thicknesses is listed below the plot, based on the results of a Wilcoxon sign-rank test.}
\end{figure}

The results of the first phase of the reproducibility analysis are summarized in \figref{fig:ccc_pairwise}.
Here, the distribution of \acp{CCC} across all feature values for each extractor is compared across each pair of slice thicknesses.
Intuitively, the largest difference in slice thickness (2.5~mm vs. 5~mm) shows a significantly lower distribution of \acp{CCC} than the other two pairs, which compare slice thicknesses which are closer together (2.5~mm vs. 3.75~mm, and 3.75~mm vs 5~mm).
These differences were highly statistically significant based on a Wilcoxon sign-rank test between all \ac{CCC} pairs ($p < 3.8\times 10^{-30}$ in the bottom two rows of the table in \figref{fig:ccc_pairwise}).
Interestingly, the 3.75~mm vs. 5~mm pairs also seem to have slightly higher \ac{CCC} than the 2.5~mm vs. 3.75~mm pairs, at a lower level of significance ($p < 6.4\times10^{-4}$ in the top row of the table in \figref{fig:ccc_pairwise}).

\begin{figure}[t]
    \centering
    \includegraphics[width=\linewidth]{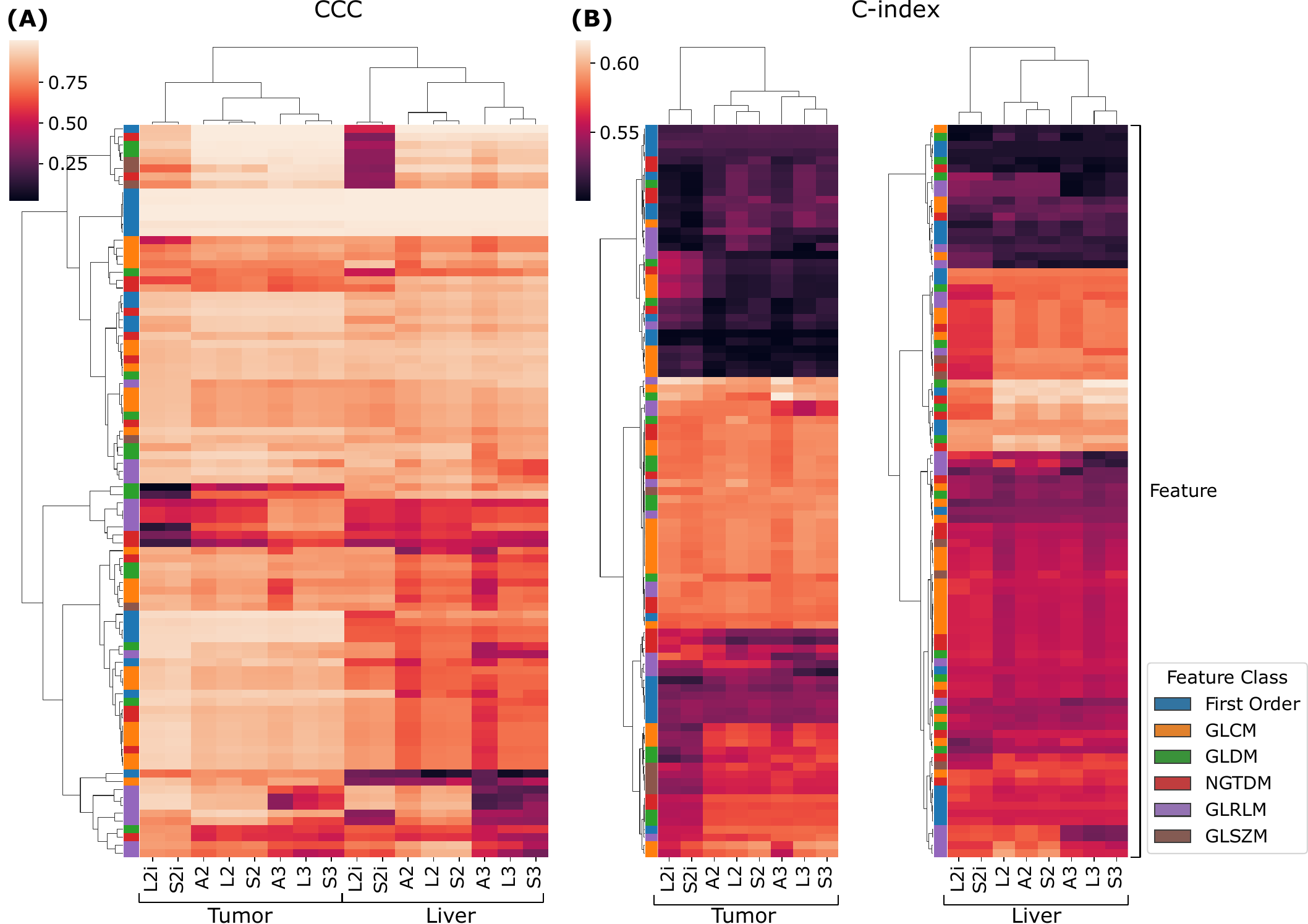}
    \caption{\label{fig:cluster}%
     Cluster maps of the \ac{CCC} (A) and \ac{C-index} (B) of all features. Each row corresponds to a unique feature, while each column is an extractor setting. For the \acp{CCC}, the liver and tumor extractor results are joined, and clustered together, to emphasize the patterns of reproducibility across \acp{ROI} and extractor. For the \ac{C-index}, the liver and tumor results are clustered separately. The feature class for each row is indicated by the left-most column of each heat map.}
\end{figure}
\begin{figure}[t]
    \centering
       \begin{tabular}{
        @{}>{\raggedright}m{0.5\linewidth}@{}
        @{}>{\raggedleft}m{0.5\linewidth}@{}
    }
        \includegraphics[width=0.98\linewidth]{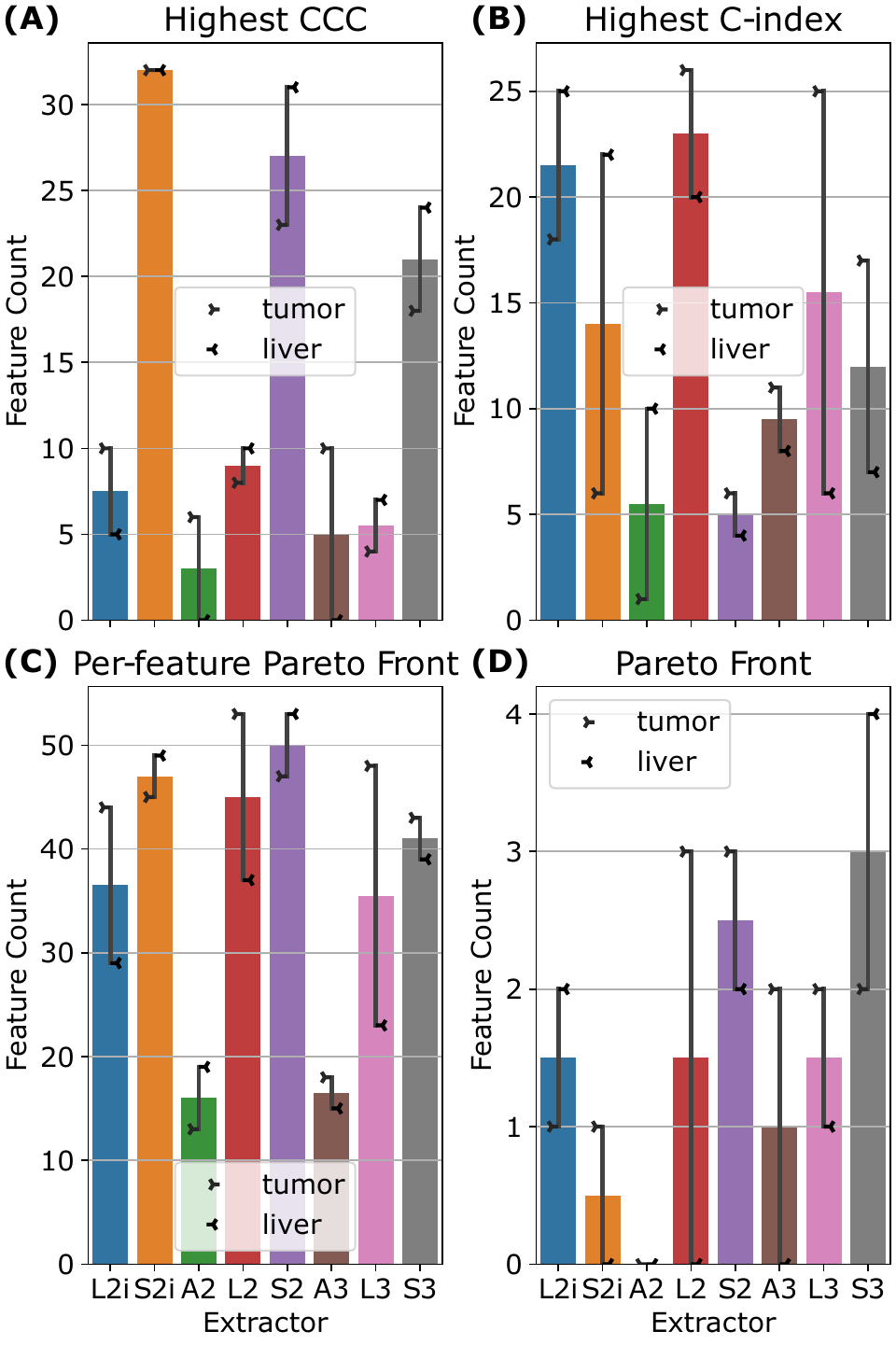} &
        \includegraphics[width=0.98\linewidth]{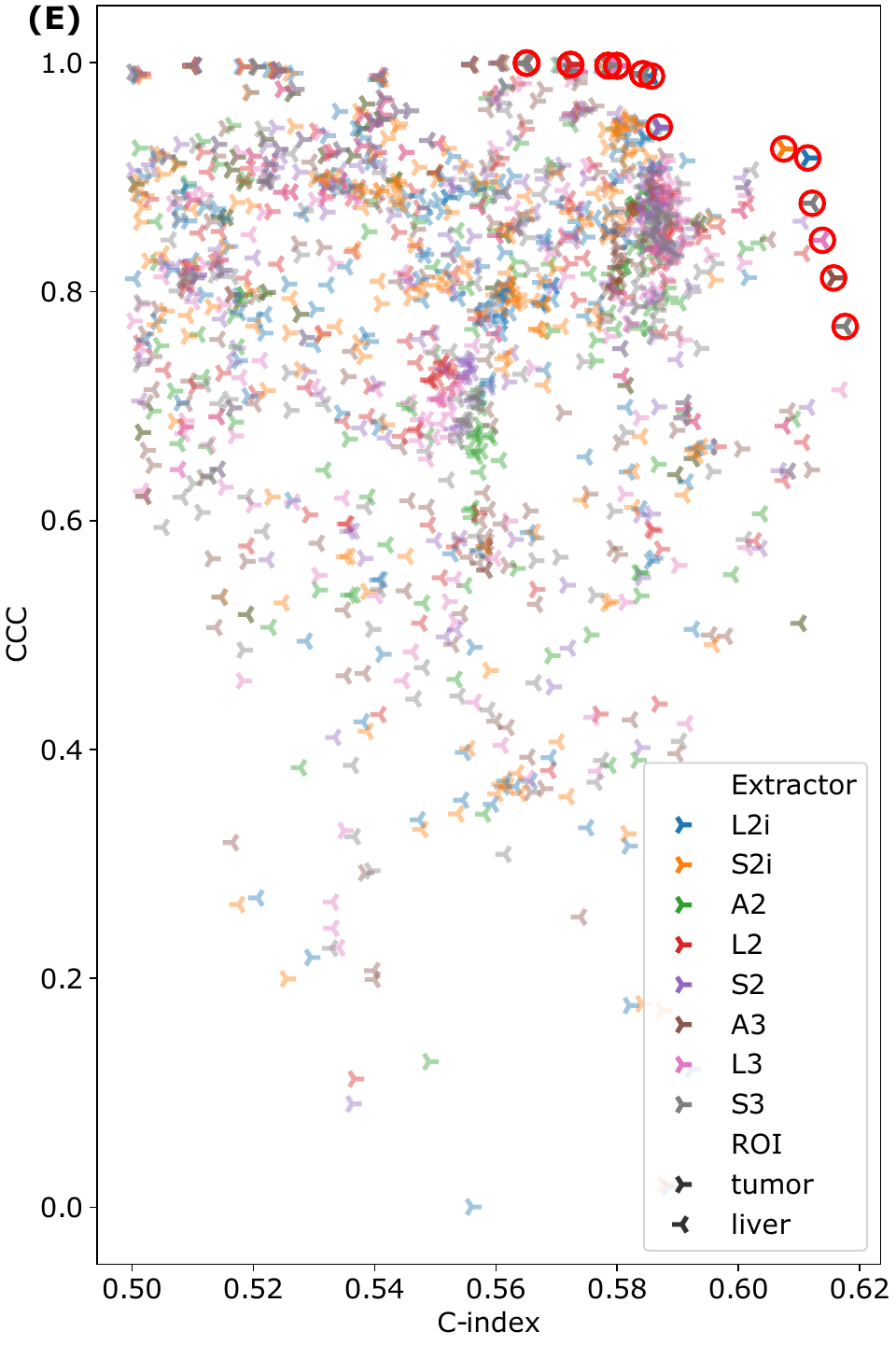}
    \end{tabular}
    
    \caption{\label{fig:pareto}%
    (A)--(C): Bar plots counting for how many features each extractor produces the highest CCC (A), highest C-index (B), or a pair of CCC and C-index that is Pareto efficient for that feature (C).
    (D): A bar plot of how many features from each extractor are on the Pareto front for \emph{all} features across all extractors. In (A)-(D), the line indicators show the number of features broken down by ROI (tumor or liver for left and right, respectively), while the bar height corresponds to the average across the two ROIs.
    (E): A scatter plot of the C-index and CCC for all features, color coded by ROI, with points on the Pareto front rendered with full opacity and circled in red.
    }
\end{figure}

We now turn to the second phase of the reproducibility analysis, based on the generalized \acp{CCC} computed across all slice thicknesses using the \ac{LMM} model.
\figref{fig:cluster} (A) summarizes the results across all features, extraction settings, and \acp{ROI}.
The results are visualized as a heat map, where each row is a feature, and each column corresponds to a combination of \ac{ROI} and extractor setting, and the values are \acp{CCC}.
Both the rows, and columns underwent hierarchical clustering based on Euclidean distance and Ward linkage.
We can see, near the top of the map, a cluster of features that are highly reproducible across all extraction settings.
As can be seen in the left-most column of the map, these highly reproducible features all belong to the first-order feature class.
With some exceptions, within both \acp{ROI}, the \acp{CCC} across extractors tend to follow similar patterns; however, overall, the liver features visually appear to tend toward lower \acp{CCC}.
This tendency is confirmed by a Wilcoxon sign-rank test, which shows that feature \acp{CCC} between the two \acp{ROI} are statistically significantly lower in the liver parenchyma, with $p \in [8.8\times 10^{-11}, 3.6\times 10^{-5}]$ across all extractors.
Interestingly, the \acp{CCC} between \acp{ROI} and extractor settings are sufficiently different that they cluster logically: the two \acp{ROI} form two higher level clusters; and within each \ac{ROI}, L2i and S2i form their own cluster, and within the other cluster, the 2.5D and 3D settings cluster together.

\figref{fig:cluster} (B) and (C) show similar cluster maps of the univariate C-indexes, computed against overall survival in the publicly available cohort, for each feature, from the tumor and liver \acp{ROI}, respectively.
The \acp{ROI} were clustered separately because the \acp{ROI} are biologically different, and therefore the discriminative ability of a given feature may be completely different in a different \ac{ROI}.
Both maps cluster features into three broad groups---a low C-index group at the top, a high C-index group in the middle, and a middle group at the bottom.
Both the high C-index and low C-index groups are larger in the tumor \ac{ROI} than in the liver parenchyma.
Broadly speaking, the features tend to show similar trends in terms of predictive value across the extractor settings.
On the other hand, there are some exceptions: at the bottom of the tumor cluster map (B), there is a cluster of features where the L2i and S2i features tend to be less predictive than the others. Similarly, in the high C-index cluster of the liver \ac{ROI}, L2i and S2i again tend to be less predictive.
The column clusters in both maps reproduce the finding for \ac{CCC} in \figref{fig:cluster} (A): three high-level clusters corresponding to (L2i and S2i), (A2, L2, and S2), and (A3, L3, and S3).

\begin{table}[t]
\caption{\label{tab:surv_top_10} The performance of the top 10 parameter combinations in terms of test-set C-index and 95\% confidence interval, in the multivariable \ac{CPH} experiments. The average proportion of features drawn from the liver across all runs is listed in the final column.}
\centering
\input{top_multivar_performers.txt}

\end{table}

\begin{figure}[t]
    \centering
    \includegraphics[width=0.55\linewidth]{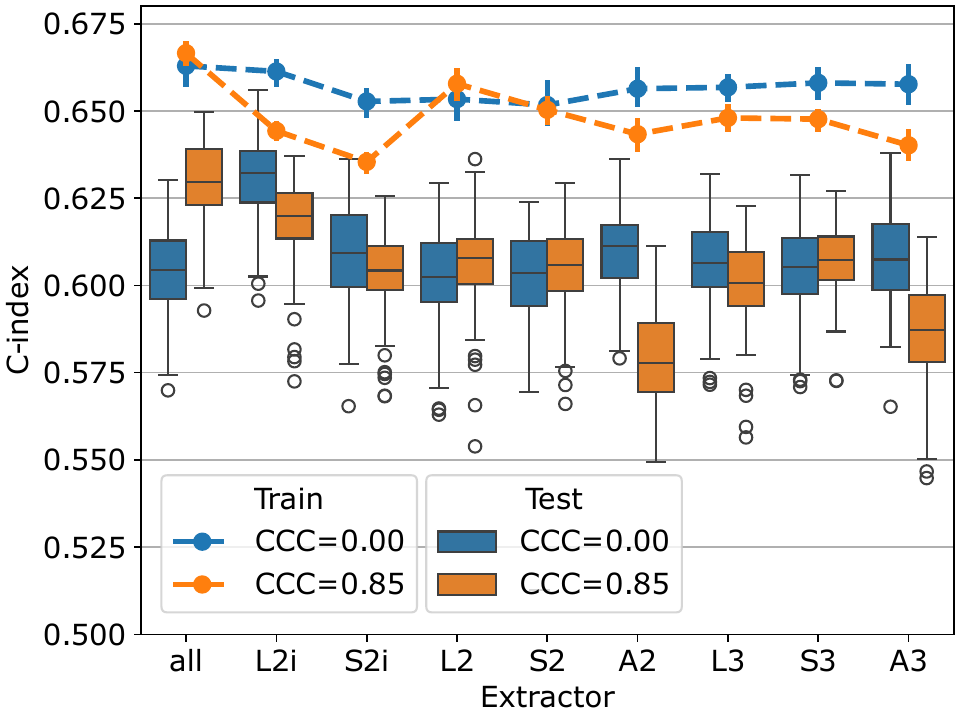}
    \caption{\label{fig:multivar_boxplot}The cross-validation performance of the 4-feature models across extractor settings is plotted for $\mathrm{CCC}_t=0$ (blue) and $\mathrm{CCC}_t = 0.85$ (orange). The test set results are summarized in box plots, while the average performance on the train set across folds is summarized by lines. The train set error bars correspond to 90\% confidence intervals for the performance across repeated cross-validations. }
\end{figure}

To examine the relationship between feature reproducibility and discriminative ability, we considered the relationships between the \ac{CCC} and \ac{C-index} for each feature.
For every feature, we checked which extractors produced the highest CCC, the highest C-index, and which extractors were Pareto efficient for both CCC and C-index in that feature.
The results are summarized in \figref{fig:pareto} (A)--(C).
Comparing \figref{fig:pareto} (A) and (B), we see that the results are inconsistent---there is no set of extractors that produce both the highest \ac{CCC} and the highest C-index.
With the exceptions of A2 and A3, \figref{fig:pareto} (C) shows that all feature extraction settings are well-represented on the Pareto fronts across features.
Finally, grouping all features from all extractors together, and computing the set of Pareto efficient features for CCC and C-index shows that all extractors except A2 contribute (see \figref{fig:pareto} (D)), with S3 contributing the greatest number of features (6).
A scatter plot of all features is shown, color coded by extractor setting, with the Pareto front highlighted, in \figref{fig:pareto} (E).
Finally, the set of all 23 Pareto efficient features is listed in \tabref{tab:pareto} in the appendix.
It is worth noting that several features that are listed are actually equal, and therefore over-counted in \figref{fig:pareto} (D).
Furthermore, note that first-order features across S2 and S3, or L2 and L3 are also equal, and therefore some features that appear in both sets are still identical.
These groups of equal features are marked in \tabref{tab:pareto}.
\figref{fig:pareto} (A)--(D) also indicates that the optimal feature extraction method for a given criterion (C-index, CCC, or Pareto front) appears to vary not only across specific features, but across \acp{ROI}.
Indeed, most of the extractors, and both \acp{ROI} are represented, at least to some extent, on the overall Pareto front across all features and extractors (see \figref{fig:pareto} (D)).

The multivariable \ac{CPH} cross-validation experiments produced results for 315 combinations of three parameters (feature extraction setting (8 extractors plus 1 for all), feature count (1, 2, 4, 8, 16, 32, 64), CCC threshold (0, 0.8, 0.85, 0.9, 0.95)).
The test-set C-index and 95\% confidence intervals across all 100 repetitions are listed for the ten highest performing parameter combinations in \tabref{tab:surv_top_10}.
Models based on the features from extractor L2i attained the highest C-index of 0.630 (0.603--0.649), with 4 features being selected, and no \ac{CCC} thresholding.
The second highest performance of 0.629 (0.605--0.645) was attained using features from all extractors, with 4 features being selected, and a threshold of $\mathrm{CCC} \geq 0.85$.
Based on these top two performing cases, the performance of the models using 4 features, across all extractors, and $\mathrm{CCC_t}=0$ and $\mathrm{CCC}_t = 0.85$ are visualized in \figref{fig:multivar_boxplot}.
For most extractors, the thresholding of features to those with CCC over 0.85 appears to have little effect on the performance of the resulting models.
In a few cases, L2i, A2, and A3, the performance is reduced after thresholding.
When including features from all extractors (left-most in the figure), the performance increases after \ac{CCC} thresholding.
The performance on the training set tends to be lower after thresholding, suggesting that the models using reproducible features may be less prone to overfitting.

\begin{figure}[!ht]
\centering
\includegraphics[width=0.95\linewidth]{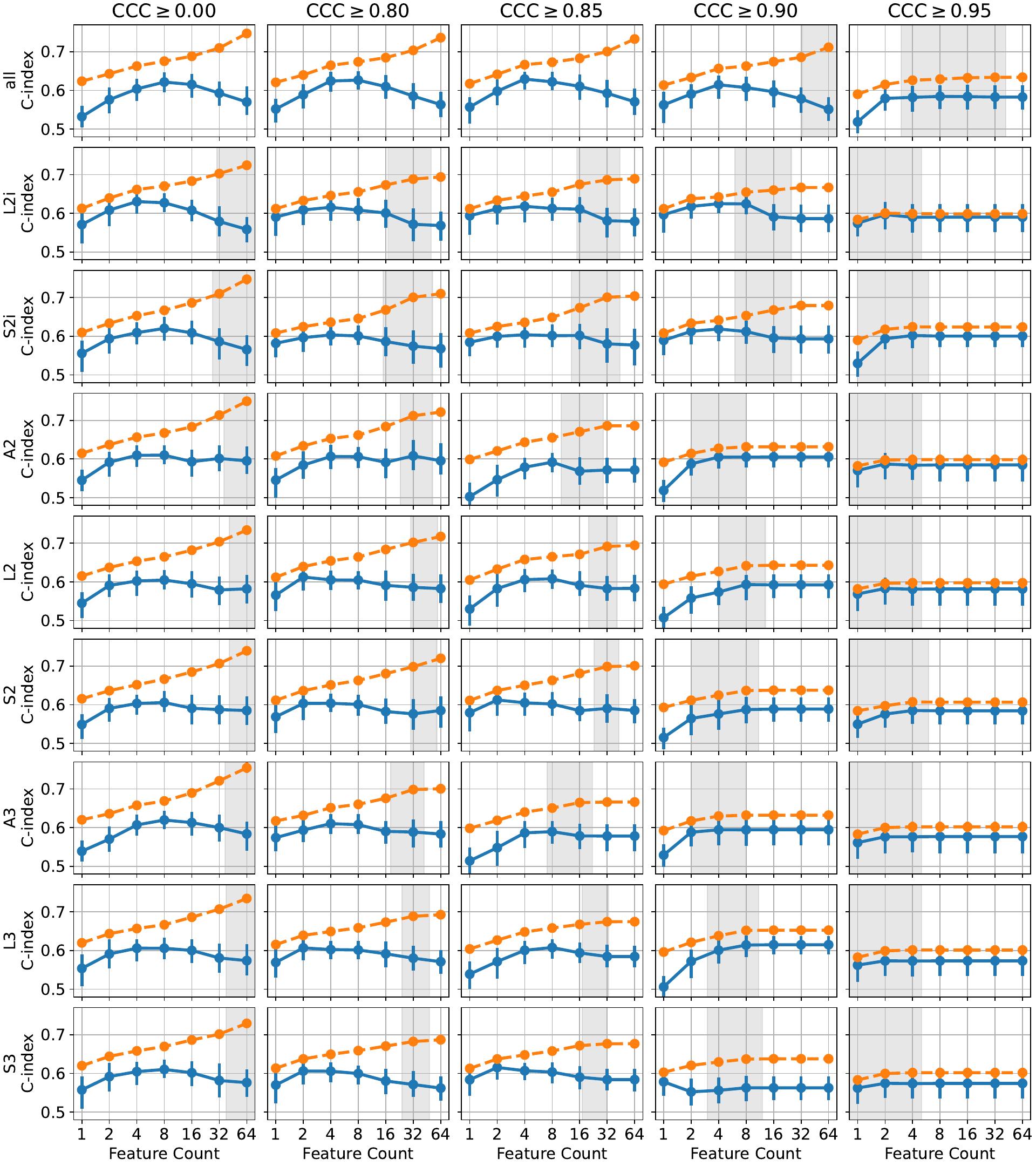}
\caption{\label{fig:multivar_sensitivity} The C-index for the train set (blue) and test set (orange) is plotted against feature count, for every extractor setting (rows) and every CCC threshold (columns). The 90\% confidence interval for the number of available features after CCC thresholding and univariate filtering is indicated by the gray regions.}
\end{figure}

\begin{figure}[t]
    \includegraphics[width=0.95\linewidth]{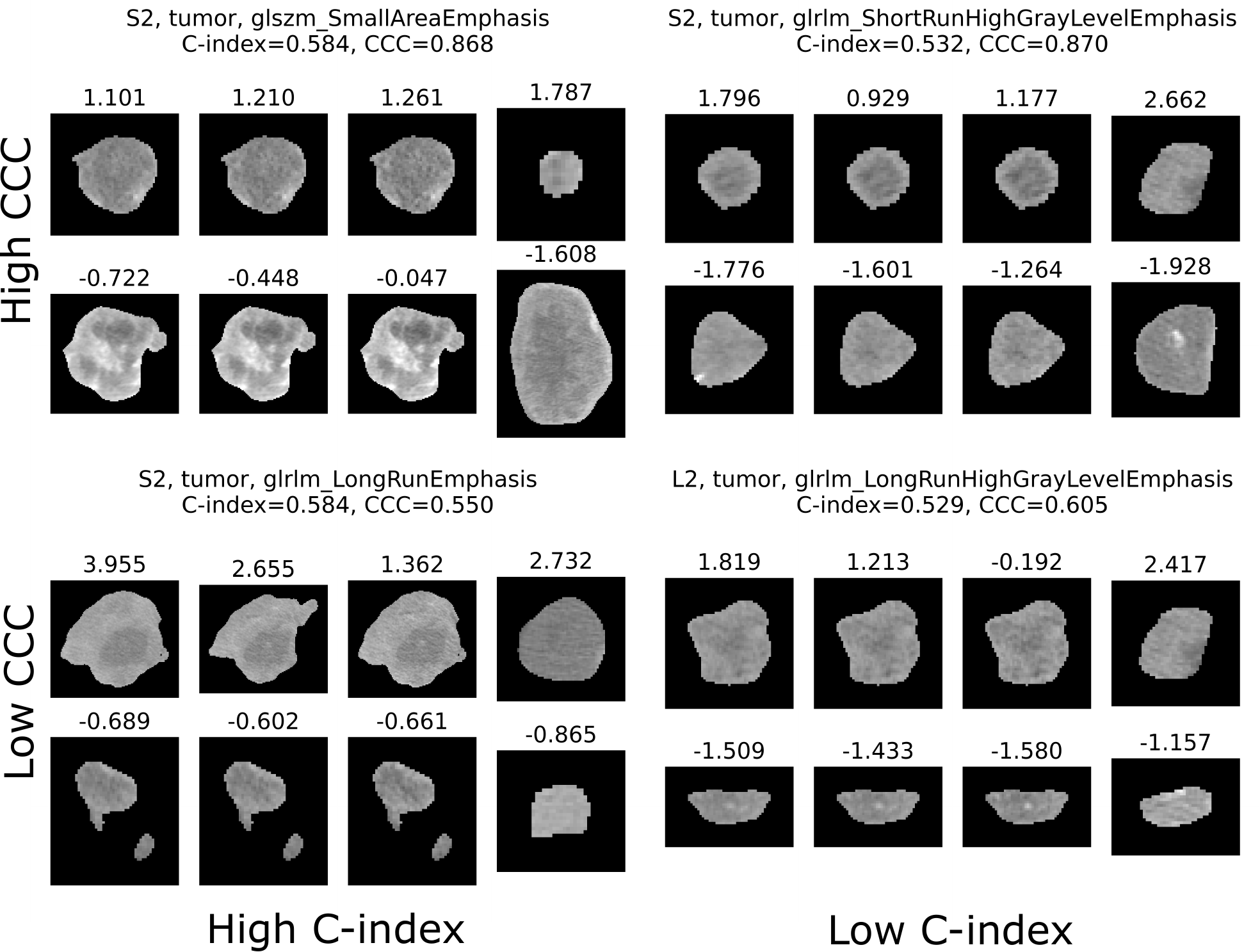}
    \caption{\label{fig:featvis} Features with high/low CCC (top/bottom) and high/low C-index (left/right) are visualized. For each, an example tumor with a high value for the selected feature, and a low value are illustrated in the two rows. The columns display the same tumor for an example patient from the reproducibility data set (first three columns, corresponding to slice thickness 5~mm, 3.75~mm and 2.5~mm), and the survival data set (fourth column).}
\end{figure}

To visualize the sensitivity of these results with respect to the \ac{CCC} threshold and the number of features selected, we have visualized the results from all parameter settings in \figref{fig:multivar_sensitivity}.
Each plot shows the C-index for both the training and test sets, plotted against the number of features selected in the model.
In some cases, after thresholding, and applying the univariate feature filter (removing features with C-index below 0.55 or with $p$-value over 0.1), there may have been fewer than the desired number of features remaining for feature selection.
In these cases, all features passing the threshold and univariate filter were used.
The 95\% confidence intervals for the number of available features at the time of applying \ac{MRMR} for the different threshold levels is indicated by the gray regions, where either limit was less than 64.
Typically, beyond this point, requesting more features does not change the performance of either the train or test set, because the number of features is saturated.
This effect is clearly visible in the plots for $\mathrm{CCC}\geq 0.9$ and $\mathrm{CCC}\geq 0.95$, where fewer features tend to remain after thresholding, due to the stringent reproducibility requirement.
Broadly, we observe that the best performance tends to be obtained using 4 or 8 features, insofar as there are sufficiently many features available after thresholding.
Again we can observe that, although the effect appears small, the training set performance is slightly reduced when the features are thresholded on $\mathrm{CCC}$, indicating a potential reduction in overfitting.
For higher thresholds, this effect is primarily due to the number of features being saturated, but for $\mathrm{CCC} \geq 0.8$ and $\mathrm{CCC}\geq 0.85$ the number of available features is high enough to observe the difference even before feature count saturation.

A visualization of tumors with different values for features with high/low C-index (prognostic value), and high/low CCC (reproducibility) is given in \figref{fig:featvis}.

\section{Discussion}

Although in the present study we have restricted our analysis to the reproducibility of radiomic features when slice thickness is varied at reconstruction time, the ultimate purpose of the data collection effort from which this study draws is to look at the effects of contrast timing and reconstruction parameters using a test-retest paradigm with two portal venous phase images collected within 15s of each other.
As part of a multicenter prospective study systematically varying contrast timing and reconstruction parameters, this paper presents early results from a research effort that will be a substantial step forward in the understanding of the reproducibility of radiomics for contrast-enhanced \ac{CT} of \ac{CRLM}.

Our results show that when comparing the 20\% ASiR images pairwise at different slice thicknesses, the CCCs for features from the 5~mm and 2.5~mm images are significantly lower than when we compare 5~mm and 3.75~mm, or 2.5 and 3.75~mm pairs.
The degradation in consistency of features as the change in slice thickness increases is consistent with findings in the literature from phantom studies~\citep{zhao2014exploring,kim2019effect,ligero2021minimizing}, as well as
lung cancer \ac{CT}~\citep{lu2016assessing,park2019deep,erdal2020quantitative}.
The consistency of this finding across cancers is noteworthy given that feature reproducibility is anatomy/disease-specific, even within \ac{CT}~\citep{vanTimmeren2016test-retest}.
We also found, at a lower level of statistical significance, that the 
features had higher CCC in the 5~mm and 3.75~mm pairs, compared to the 2.5~mm and 3.75~mm pairs.
We hypothesize that this result could be due to the lower noise level in thicker sliced images, compared to thinner slices, or due to resampling issues when upsampling the \ac{ROI} segmentations to 2.5~mm.

Moving to liver cancer more specifically, \cite{perrin2018short-term} examined a database of consecutive patients from \ac{MSK} diagnosed with liver malignancy, with the additional inclusion requirement that they had two contrast-enhanced abdominal \ac{CT} scans taken within no more than 14 days of each other. 
By including two scans per patient with a small separation in time, this data set served as an approximation of a test-retest study, allowing the study of the reproducibility of radiomic features across two consecutive scans.
The database included patients with multiple liver cancers including liver metastases (n=22), intrahepatic cholangiocarcinoma (n=10), and hepatocellular carcinoma (n=6). 
Because image acquisition and reconstruction parameters naturally varied between the scans for a given patient, this data set allowed the exploration of the effects of these different parameters on the overall concordance of the radiomic features between the consecutive scans, as measured by \ac{CCC}.
One parameter that \cite{perrin2018short-term} reported on was pixel spacing, which is closely related to slice thickness, given that both variables directly affect the resulting voxel volume of the 3D image.
They found that scan pairs with a greater difference in pixel spacing had lower feature agreement for features extracted from both tumor and liver parenchyma \ac{ROI}s, highlighting along with our own results that the feature reproducibility across variations of any parameters affecting voxel volume, including slice thickness and field of view, is an important consideration when drawing inference from radiomic features.
For all the variables they considered, including pixel spacing, \cite{perrin2018short-term} also found that the tumor tended to have a larger number of reproducible features than the liver parenchyma.
Similarly, our reproducibility analysis showed that features extracted from the liver parenchyma were overall less reproducible across slice thickness variation than those drawn from the largest metastasis, regardless of the extractor setting under consideration.

Despite the lower reproducibility of the features drawn from the liver parenchyma, our univariate survival analysis showed that the liver features still contained a relevant signal with regard to overall patient survival in the independent retrospective data set.
Indeed, the highest overall univariate C-index was attained by a feature from the liver parenchyma---GLDM SmallDependenceLowGrayLevelEmphasis (C-index=0.6176)---and features from the liver parenchyma were well-represented on the Pareto front for C-index and CCC (see \tabref{tab:pareto}).
Similarly, in the top-performing multivariable models, across all cross-validation runs, the average proportion of selected features coming from the liver was usually over 50\% (see \tabref{tab:surv_top_10}).
The presence of important information relevant to patient outcomes in the radiological texture of the liver parenchyma is biologically feasible, and consistent with previous analyses in the literature~\citep{simpson2017computed,hu2022radiomics}.
Given the apparent importance of the liver parenchyma features, future studies should investigate methods of tailoring the feature extraction to improve the reproducibility of the features extracted from the liver parenchyma, without degrading their prognostic value.

The \ac{IBSI} reference manual~\citep{zwanenburg2016image} recommends resampling to a common voxel size to ensure the features are comparable between images with different resolutions.
However, several empirical studies have shown that resampling has a limited ability to mitigate the effects of slice thickness, or resolution generally, on radiomic features.
\cite{ligero2021minimizing} showed in a phantom study that resampling could improve feature concordance across varying slice thickness to a limited extent, but found that harmonization of features treating slice thickness as a batch effect had a stronger effect.
In lung \ac{CT}, studies have found limited improvement when resampling images to the same voxel volume with linear interpolation~\citep{yang2021evaluation}, although deep-learning-based super-resolution algorithms have shown promise~\citep{park2019deep}.
In a study varying pixel spacing by varying field of view during reconstruction, \cite{mackin2017harmonizing} found that resampling on its own actually worsened reproducibility of features across pixel sizes; however, combining resampling with a low-pass Butterworth filter improved feature concordance.
\cite{shafiq-ul-hassan2017intrinsic} found in a phantom study that resampling improved feature concordance across images acquired with different voxel sizes for only a small subset of features.
The majority of these features could also be corrected without resampling, by including a correction factor based on the voxel volume and/or \ac{ROI} volume---a result which they subsequently validated in a lung cancer cohort~\citep{shafiq-ul-hassan2018voxel}.
Using a similar approach to feature correction, \cite{escuderoSanchez2021robustness}, in a study of contrast enhanced \ac{CT} imaging of liver cancer, found that while resampling to the mean spacing in all directions improved the number of features that were reproducible across changes in slice thickness, after feature correction, the result was modest in terms of absolute improvement over extracting features from the original images.
In aggregate, these studies show that isotropic resampling is, on its own, not sufficient to mitigate the effects of slice thickness variation on the resulting radiomic feature values.

Our results also indicate limitations of resampling to mitigate voxel resolution effects.
In the present study, we were further motivated to question the use of isotropic resampling---as recommended by \ac{IBSI}~\citep{zwanenburg2016image} to preserve rotational invariance of feature definitions---due to the inherently anisotropic nature of our abdominal images.
In our previous study on this topic~\citep{peoples2023examining}, we showed that aggregating features in 2.5D with no z-axis resampling (i.e. L2i and S2i) tended to produce more reproducible texture features, and still produced good prognostic models.
In the present work, we conducted a more in-depth analysis of the relationship between feature extraction method, reproducibility, and prognostic value.
From \figref{fig:cluster}, we can see that no one extraction method produces more reproducible or more prognostic features in every case.
Indeed, both \figref{fig:cluster} and \figref{fig:pareto} illustrate that there is no consistently ``best'' feature extraction approach across all features and \acp{ROI}, when we consider either C-index or CCC in isolation, or when we consider both by looking at the Pareto front.
Furthermore, while the best overall performance in the multivariable survival models was obtained by using features from L2i, the combination of all extractors was able to rival this performance when we removed features with $\mathrm{CCC} < 0.85$ across slice thicknesses.
These results suggest that when feature reproducibility across important variations in imaging protocol is known a priori (such as from a test-retest study), including reproducibility into the feature selection process can improve model performance.
In this case, highly reproducible and prognostic models can be achieved without optimizing the feature extraction process, by including features extracted using a variety of settings, and allowing the best features to be selected in a data-driven manner accounting for both reproducibility and discriminative ability.
This data-driven approach is reminiscent of a method used by \cite{vallieres2017radiomics}, wherein features extracted with multiple settings, such as resampling resolutions, were pooled into a larger table, and allowed to be considered as separate features during model building.
In the absence of reproducibility scores for each feature, choosing settings that can produce features that are more robust to protocol variations present in the data set may be more important, but it is unclear how to choose these settings without a reproducibility study, given the study-specific nature of feature \acp{CCC}~\citep{vanTimmeren2016test-retest}.

One limitation of this study is the use of only one fixed bin count of 24 when computing the texture features.
The bin count is a key parameter for computing texture features, and has a large effect on the results;
though, for many features this effect is at least somewhat predictable~\citep{shafiq-ul-hassan2017intrinsic,shafiq-ul-hassan2018voxel}.
Furthermore, phantom studies have suggested that variation of the discretization, although it affects the feature values, does not have a large effect on feature reproducibility~\citep{larue2017influence}.
Due to the large number of modeling experiments, and already large number of dimensions under consideration in the present study, we chose to avoid adding bin count as an additional variable at this time.
In a study of contrast enhanced \ac{CT} of hepatocellular carcinoma, \cite{escuderoSanchez2021robustness} found that the optimal bin count for feature reproducibility across slice thicknesses was in the range of (32--64), which given the step-size in their analysis, is nearly encompassing our chosen value of 24.
Depite this, future work should consider the effect of bin count on feature reproducibility and prognostic value.

In conclusion, our results demonstrate the strong effect of slice thickness on feature reproducibility for contrast enhanced \ac{CT} imaging of \ac{CRLM}.
Although some methods of feature extraction may mitigate the effects of slice thickness on some features, overall, we found that the extractor producing the most reproducible value for a given feature can vary across features and \acp{ROI}.
Similarly, the greatest discriminative ability for a given feature may be attained by different extractors, dependent on feature, and ROI.
Given this, our results support a data-driven approach, where features from a variety of extractor settings are all considered, and selected in a manner accounting for reproducibility across relevant variations in protocol.
Where disease-specific reproducibility metrics are not available, some methods of feature extraction may perform better due to improvements in reproducibility across certain prognostic features (such as the L2i features in the present study), however it is unclear how to determine this without a reproducibility study.
Overall, our results demonstrate that we can find radiomic features that are both reproducible across slice thickness variation, and prognostic in patients undergoing hepatic resection, in the context of contrast enhanced \ac{CT} of colorectal liver metastases.

\acks{This work was supported by National Institutes of Health grant R01 CA233888.}

\ethics{The work follows appropriate ethical standards in conducting research and writing the manuscript, following all applicable laws and regulations regarding treatment of animals or human subjects.
The study was conducted in accordance with the principles described in the declaration of Helsinki, with approval of our local IRB and informed consent of all patients.
}

\coi{We declare that we have no conflicts of interest.}

\data{The retrospective data set used for survival analysis is publicly available from \ac{TCIA}~\citep{simpson2023preoperative}. The prospective data set used for the reproducibility analysis is still being actively collected and prepared, and is therefore not available, though our intent is ultimately to release it on \ac{TCIA}.}

\renewcommand{\doi}[1]{doi: \url{#1}}
\bibliography{latex/paperpile_citations}

\clearpage
\appendix
\section{Additional Results}
\tabref{tab:pareto} lists all features on the Pareto front for \ac{CCC} and \ac{C-index} when combining all feature extractor settings and \acp{ROI}.

\begin{table}[h]
\caption{\label{tab:pareto} Features on the Pareto front for \ac{CCC} and \ac{C-index} when combining all feature extractor settings and \acp{ROI}.}
\input{pareto_features.txt}

\end{table}

\tabref{tab:feature_list} lists all features used in this study.
\input{list_all_features.txt}

\end{document}

%% file: feature_counts_by_class.txt
\begin{tabular}{lr}
\toprule
Feature Class & Count \\
\midrule
First order & 18 \\
GLCM & 24 \\
NGTDM & 5 \\
GLDM & 14 \\
GLRLM & 16 \\
GLSZM & 16 \\ \addlinespace
\bfseries Total & \bfseries 93 \\
\bottomrule
\end{tabular}

%% file: top_multivar_performers.txt
\begin{tabular}{lrrlr}
\toprule
Extractor & Feature Count & $\mathrm{CCC}_t$ & Harrel's C-index & Prop. Liver Features \\
\midrule
L2i & 4 & 0.00 & 0.630 (0.603--0.649) & 0.52 \\
all & 4 & 0.85 & 0.629 (0.605--0.645) & 0.53 \\
L2i & 8 & 0.00 & 0.627 (0.607--0.648) & 0.57 \\
all & 8 & 0.80 & 0.626 (0.606--0.645) & 0.55 \\
L2i & 4 & 0.90 & 0.625 (0.603--0.643) & 0.28 \\
all & 4 & 0.80 & 0.624 (0.600--0.645) & 0.53 \\
L2i & 8 & 0.90 & 0.624 (0.602--0.645) & 0.24 \\
all & 8 & 0.85 & 0.622 (0.605--0.644) & 0.56 \\
all & 8 & 0.00 & 0.622 (0.599--0.643) & 0.50 \\
S2i & 8 & 0.00 & 0.620 (0.592--0.645) & 0.61 \\
\bottomrule
\end{tabular}

%% file: pareto_features.txt
\begin{tabular}{llllrr}
\toprule
Extractor & ROI &  Class & Name & CCC & C-index \\
\midrule
\multirow[t]{3}{*}{L2i} & Tumor & GLSZM & ZoneEntropy & 0.9169 & 0.6114 \\
 & Liver & First order & 10Percentile & 0.9885 & 0.5857 \\
 & Liver & First order & Mean & 0.9998 & 0.5650 \\
S2i & Tumor & GLSZM & ZoneEntropy & 0.9248 & 0.6075 \\
\multirow[t]{3}{*}{L2} & Tumor & First order & Energy & 0.9974 & 0.5800 \\
 & Tumor & First order & TotalEnergy & 0.9974 & 0.5800 \\
 & Tumor & GLRLM & GrayLevelNonUniformity & 0.9986 & 0.5723 \\
\multirow[t]{5}{*}{S2} & Tumor & First order & Energy & 0.9976 & 0.5785 \\
 & Tumor & First order & TotalEnergy & 0.9976 & 0.5785 \\
 & Tumor & GLDM & SmallDependenceHighGrayLevelEmphasis & 0.9439 & 0.5869 \\
 & Liver & First order & 10Percentile & 0.9904 & 0.5842 \\
 & Liver & First order & Mean & 0.9997 & 0.5651 \\
\multirow[t]{2}{*}{A3} & Tumor & GLDM & DependenceEntropy & 0.8120 & 0.6157 \\
 & Tumor & GLRLM & GrayLevelNonUniformity & 0.9985 & 0.5724 \\
\multirow[t]{3}{*}{L3} & Tumor & First order & Energy & 0.9974 & 0.5800 \\
 & Tumor & First order & TotalEnergy & 0.9974 & 0.5800 \\
 & Liver & GLRLM & ShortRunLowGrayLevelEmphasis & 0.8450 & 0.6138 \\
\multirow[t]{6}{*}{S3} & Tumor & First order & Energy & 0.9976 & 0.5785 \\
 & Tumor & First order & TotalEnergy & 0.9976 & 0.5785 \\
 & Liver & First order & 10Percentile & 0.9904 & 0.5842 \\
 & Liver & First order & Mean & 0.9997 & 0.5651 \\
 & Liver & GLDM & SmallDependenceLowGrayLevelEmphasis & 0.7696 & 0.6176 \\
 & Liver & GLRLM & ShortRunLowGrayLevelEmphasis & 0.8773 & 0.6122 \\
\bottomrule
\end{tabular}

%% file: list_all_features.txt
\begin{longtable}{ll}
\caption{All features from all feature classes. Detailed feature definitions can be found in the \texttt{pyradiomics} documentation. \url{https://pyradiomics.readthedocs.io/en/latest/features.html}} \label{tab:feature_list} \\
\toprule
Feature Class & Name \\
\midrule
\endfirsthead
\caption[]{All features from all feature classes.} \\
\toprule
Feature Class & Name \\
\midrule
\endhead
\midrule
\multicolumn{2}{r}{Continued on next page} \\
\midrule
\endfoot
\bottomrule
\endlastfoot
\multirow[t]{18}{*}{First order} & 10Percentile \\
 & 90Percentile \\
 & Energy \\
 & Entropy \\
 & InterquartileRange \\
 & Kurtosis \\
 & Maximum \\
 & Mean \\
 & MeanAbsoluteDeviation \\
 & Median \\
 & Minimum \\
 & Range \\
 & RobustMeanAbsoluteDeviation \\
 & RootMeanSquared \\
 & Skewness \\
 & TotalEnergy \\
 & Uniformity \\
 & Variance \\
\cline{1-2}
\multirow[t]{24}{*}{GLCM} & Autocorrelation \\
 & ClusterProminence \\
 & ClusterShade \\
 & ClusterTendency \\
 & Contrast \\
 & Correlation \\
 & DifferenceAverage \\
 & DifferenceEntropy \\
 & DifferenceVariance \\
 & Id \\
 & Idm \\
 & Idmn \\
 & Idn \\
 & Imc1 \\
 & Imc2 \\
 & InverseVariance \\
 & JointAverage \\
 & JointEnergy \\
 & JointEntropy \\
 & MCC \\
 & MaximumProbability \\
 & SumAverage \\
 & SumEntropy \\
 & SumSquares \\
\cline{1-2}
\multirow[t]{14}{*}{GLDM} & DependenceEntropy \\
 & DependenceNonUniformity \\
 & DependenceNonUniformityNormalized \\
 & DependenceVariance \\
 & GrayLevelNonUniformity \\
 & GrayLevelVariance \\
 & HighGrayLevelEmphasis \\
 & LargeDependenceEmphasis \\
 & LargeDependenceHighGrayLevelEmphasis \\
 & LargeDependenceLowGrayLevelEmphasis \\
 & LowGrayLevelEmphasis \\
 & SmallDependenceEmphasis \\
 & SmallDependenceHighGrayLevelEmphasis \\
 & SmallDependenceLowGrayLevelEmphasis \\
\cline{1-2}
\multirow[t]{16}{*}{GLRLM} & GrayLevelNonUniformity \\
 & GrayLevelNonUniformityNormalized \\
 & GrayLevelVariance \\
 & HighGrayLevelRunEmphasis \\
 & LongRunEmphasis \\
 & LongRunHighGrayLevelEmphasis \\
 & LongRunLowGrayLevelEmphasis \\
 & LowGrayLevelRunEmphasis \\
 & RunEntropy \\
 & RunLengthNonUniformity \\
 & RunLengthNonUniformityNormalized \\
 & RunPercentage \\
 & RunVariance \\
 & ShortRunEmphasis \\
 & ShortRunHighGrayLevelEmphasis \\
 & ShortRunLowGrayLevelEmphasis \\
\cline{1-2}
\multirow[t]{16}{*}{GLSZM} & GrayLevelNonUniformity \\
 & GrayLevelNonUniformityNormalized \\
 & GrayLevelVariance \\
 & HighGrayLevelZoneEmphasis \\
 & LargeAreaEmphasis \\
 & LargeAreaHighGrayLevelEmphasis \\
 & LargeAreaLowGrayLevelEmphasis \\
 & LowGrayLevelZoneEmphasis \\
 & SizeZoneNonUniformity \\
 & SizeZoneNonUniformityNormalized \\
 & SmallAreaEmphasis \\
 & SmallAreaHighGrayLevelEmphasis \\
 & SmallAreaLowGrayLevelEmphasis \\
 & ZoneEntropy \\
 & ZonePercentage \\
 & ZoneVariance \\
\cline{1-2}
\multirow[t]{5}{*}{NGTDM} & Busyness \\
 & Coarseness \\
 & Complexity \\
 & Contrast \\
 & Strength \\
\cline{1-2}
\end{longtable}

%% file: unsure2023_melba.bbl
\begin{thebibliography}{54}
\providecommand{\natexlab}[1]{#1}
\providecommand{\url}[1]{\texttt{#1}}
\expandafter\ifx\csname urlstyle\endcsname\relax
  \providecommand{\doi}[1]{doi: #1}\else
  \providecommand{\doi}{doi: \begingroup \urlstyle{rm}\Url}\fi

\bibitem[Bar-Joseph et~al.(2001)Bar-Joseph, Gifford, and Jaakkola]{bar-joseph2001optimal}
Ziv Bar-Joseph, David~K Gifford, and Tommi~S Jaakkola.
\newblock Fast optimal leaf ordering for hierarchical clustering.
\newblock \emph{Bioinformatics}, 17\penalty0 (suppl\_1):\penalty0 S22--S29, June 2001.

\bibitem[Berenguer et~al.(2018)Berenguer, Pastor-Juan, Canales-Vázquez, Castro-García, Villas, Mansilla~Legorburo, and Sabater]{berenguer2018radiomics}
Roberto Berenguer, María Del~Rosario Pastor-Juan, Jesús Canales-Vázquez, Miguel Castro-García, María~Victoria Villas, Francisco Mansilla~Legorburo, and Sebastià Sabater.
\newblock Radiomics of {CT} features may be nonreproducible and redundant: Influence of {CT} acquisition parameters.
\newblock \emph{Radiology}, 288\penalty0 (2):\penalty0 407--415, August 2018.

\bibitem[Carrasco and Jover(2003)]{carrasco2003estimating}
Josep~L Carrasco and Lluís Jover.
\newblock Estimating the generalized concordance correlation coefficient through variance components.
\newblock \emph{Biometrics}, 59\penalty0 (4):\penalty0 849--858, December 2003.

\bibitem[Clark et~al.(2013)Clark, Vendt, Smith, Freymann, Kirby, Koppel, Moore, Phillips, Maffitt, Pringle, Tarbox, and Prior]{clark2013cancer}
Kenneth Clark, Bruce Vendt, Kirk Smith, John Freymann, Justin Kirby, Paul Koppel, Stephen Moore, Stanley Phillips, David Maffitt, Michael Pringle, Lawrence Tarbox, and Fred Prior.
\newblock The cancer imaging archive ({TCIA}): Maintaining and operating a public information repository.
\newblock \emph{J. Digit. Imaging}, 26\penalty0 (6):\penalty0 1045--1057, 2013.

\bibitem[Ding and Peng(2005)]{ding2005minimum}
Chris Ding and Hanchuan Peng.
\newblock {MINIMUM} {REDUNDANCY} {FEATURE} {SELECTION} {FROM} {MICROARRAY} {GENE} {EXPRESSION} {DATA}.
\newblock \emph{J. Bioinform. Comput. Biol.}, 03\penalty0 (02):\penalty0 185--205, 2005.

\bibitem[Emaminejad et~al.(2021)Emaminejad, Wahi-Anwar, Kim, Hsu, Brown, and McNitt-Gray]{emaminejad2021reproducibility}
Nastaran Emaminejad, Muhammad~Wasil Wahi-Anwar, Grace Hyun~J Kim, William Hsu, Matthew Brown, and Michael McNitt-Gray.
\newblock Reproducibility of lung nodule radiomic features: Multivariable and univariable investigations that account for interactions between {CT} acquisition and reconstruction parameters.
\newblock \emph{Med. Phys.}, 48\penalty0 (6):\penalty0 2906--2919, June 2021.

\bibitem[Erdal et~al.(2020)Erdal, Demirer, Little, Amadi, Ibrahim, O'Donnell, Grimmer, Gupta, Prevedello, and White]{erdal2020quantitative}
Barbaros~S Erdal, Mutlu Demirer, Kevin~J Little, Chiemezie~C Amadi, Gehan F~M Ibrahim, Thomas~P O'Donnell, Rainer Grimmer, Vikash Gupta, Luciano~M Prevedello, and Richard~D White.
\newblock Are quantitative features of lung nodules reproducible at different {CT} acquisition and reconstruction parameters?
\newblock \emph{PLoS One}, 15\penalty0 (10):\penalty0 e0240184, October 2020.

\bibitem[Escudero~Sanchez et~al.(2021)Escudero~Sanchez, Rundo, Gill, Hoare, Mendes~Serrao, and Sala]{escuderoSanchez2021robustness}
Lorena Escudero~Sanchez, Leonardo Rundo, Andrew~B Gill, Matthew Hoare, Eva Mendes~Serrao, and Evis Sala.
\newblock Robustness of radiomic features in {CT} images with different slice thickness, comparing liver tumour and muscle.
\newblock \emph{Sci. Rep.}, 11\penalty0 (1):\penalty0 8262, April 2021.

\bibitem[Fiz et~al.(2020)Fiz, Viganò, Gennaro, Costa, La~Bella, Boichuk, Cavinato, Sollini, Politi, Chiti, and Torzilli]{fiz2020radiomics}
Francesco Fiz, Luca Viganò, Nicolò Gennaro, Guido Costa, Ludovico La~Bella, Alexandra Boichuk, Lara Cavinato, Martina Sollini, Letterio~S Politi, Arturo Chiti, and Guido Torzilli.
\newblock Radiomics of liver metastases: A systematic review.
\newblock \emph{Cancers}, 12\penalty0 (10):\penalty0 2881, October 2020.

\bibitem[Ger et~al.(2018)Ger, Zhou, Chi, Lee, Layman, Jones, Goff, Fuller, Howell, Li, Stafford, Court, and Mackin]{ger2018comprehensive}
Rachel~B Ger, Shouhao Zhou, Pai-Chun~Melinda Chi, Hannah~J Lee, Rick~R Layman, A~Kyle Jones, David~L Goff, Clifton~D Fuller, Rebecca~M Howell, Heng Li, R~Jason Stafford, Laurence~E Court, and Dennis~S Mackin.
\newblock Comprehensive investigation on controlling for {CT} imaging variabilities in radiomics studies.
\newblock \emph{Sci. Rep.}, 8\penalty0 (1):\penalty0 13047, August 2018.

\bibitem[Gillies et~al.(2016)Gillies, Kinahan, and Hricak]{gillies2016radiomics}
Robert~J Gillies, Paul~E Kinahan, and Hedvig Hricak.
\newblock Radiomics: Images are more than pictures, they are data.
\newblock \emph{Radiology}, 278\penalty0 (2):\penalty0 563--577, February 2016.
\newblock ppublish.

\bibitem[Hamghalam et~al.(2021)Hamghalam, Frangi, Lei, and Simpson]{hamghalam2021modality}
Mohammad Hamghalam, Alejandro~F Frangi, Baiying Lei, and Amber~L Simpson.
\newblock Modality completion via gaussian process prior variational autoencoders for multi-modal glioma segmentation.
\newblock In \emph{Medical Image Computing and Computer Assisted Intervention--MICCAI 2021: Part VII}, pages 442--452, 2021.

\bibitem[Hamghalam et~al.(2023)Hamghalam, Do, and Simpson]{hamghalam2023attention-based}
Mohammad Hamghalam, Richard K~G Do, and Amber~L Simpson.
\newblock Attention-based {CT} scan interpolation for lesion segmentation of colorectal liver metastases.
\newblock In \emph{Proc. SPIE 12468, Medical Imaging 2023: Biomedical Applications in Molecular, Structural, and Functional Imaging}, volume 12468, pages 186--193, 2023.

\bibitem[Harrell et~al.(1996)Harrell, Lee, and Mark]{harrell1996multivariable}
F~E Harrell, Jr, K~L Lee, and D~B Mark.
\newblock Multivariable prognostic models: issues in developing models, evaluating assumptions and adequacy, and measuring and reducing errors.
\newblock \emph{Stat. Med.}, 15\penalty0 (4):\penalty0 361--387, February 1996.
\newblock ppublish.

\bibitem[He et~al.(2016)He, Huang, Ma, Liang, Liang, and Liu]{he2016effects}
Lan He, Yanqi Huang, Zelan Ma, Cuishan Liang, Changhong Liang, and Zaiyi Liu.
\newblock Effects of contrast-enhancement, reconstruction slice thickness and convolution kernel on the diagnostic performance of radiomics signature in solitary pulmonary nodule.
\newblock \emph{Sci. Rep.}, 6\penalty0 (1):\penalty0 34921, October 2016.

\bibitem[Horvat et~al.(2022)Horvat, Miranda, El~Homsi, Peoples, Long, Simpson, and Do]{horvat2022primer}
Natally Horvat, Joao Miranda, Maria El~Homsi, Jacob~J Peoples, Niamh~M Long, Amber~L Simpson, and Richard K~G Do.
\newblock A primer on texture analysis in abdominal radiology.
\newblock \emph{Abdom. Radiol. (NY)}, 47\penalty0 (9):\penalty0 2972--2985, September 2022.

\bibitem[Hu et~al.(2022{\natexlab{a}})Hu, Chen, Zhong, Lin, Yu, Hu, Tao, Lin, Niu, Chen, Wu, and Sun]{hu2022effects}
Peng Hu, Liye Chen, Yaoying Zhong, Yudong Lin, Xiaojing Yu, Xi~Hu, Xinwei Tao, Shushen Lin, Tianye Niu, Ran Chen, Xia Wu, and Jihong Sun.
\newblock Effects of slice thickness on {CT} radiomics features and models for staging liver fibrosis caused by chronic liver disease.
\newblock \emph{Jpn. J. Radiol.}, 40\penalty0 (10):\penalty0 1061--1068, October 2022{\natexlab{a}}.

\bibitem[Hu et~al.(2022{\natexlab{b}})Hu, Chen, Peoples, Salameh, Gönen, Romesser, Simpson, and Reyngold]{hu2022radiomics}
Ricky Hu, Ishita Chen, Jacob Peoples, Jean-Paul Salameh, Mithat Gönen, Paul~B Romesser, Amber~L Simpson, and Marsha Reyngold.
\newblock Radiomics artificial intelligence modelling for prediction of local control for colorectal liver metastases treated with radiotherapy.
\newblock \emph{Phys. Imaging Radiat. Oncol.}, 24:\penalty0 36--42, October 2022{\natexlab{b}}.

\bibitem[Ibrahim et~al.(2022)Ibrahim, Barufaldi, Refaee, Silva~Filho, Acciavatti, Salahuddin, Hustinx, Mottaghy, Maidment, and Lambin]{ibrahim2022maaspenn}
Abdalla Ibrahim, Bruno Barufaldi, Turkey Refaee, Telmo~M Silva~Filho, Raymond~J Acciavatti, Zohaib Salahuddin, Roland Hustinx, Felix~M Mottaghy, Andrew D~A Maidment, and Philippe Lambin.
\newblock {MaasPenn} radiomics reproducibility score: A novel quantitative measure for evaluating the reproducibility of {CT}-based handcrafted radiomic features.
\newblock \emph{Cancers}, 14\penalty0 (7):\penalty0 1599, March 2022.

\bibitem[Isensee et~al.(2021)Isensee, Jaeger, Kohl, Petersen, and Maier-Hein]{isensee2021nnu-net}
Fabian Isensee, Paul~F Jaeger, Simon A~A Kohl, Jens Petersen, and Klaus~H Maier-Hein.
\newblock {nnU}-{Net}: a self-configuring method for deep learning-based biomedical image segmentation.
\newblock \emph{Nat. Methods}, 18\penalty0 (2):\penalty0 203--211, February 2021.

\bibitem[Kikinis et~al.(2013)Kikinis, Pieper, and Vosburgh]{kikinis20133d}
Ron Kikinis, Steve~D Pieper, and Kirby~G Vosburgh.
\newblock {3D Slicer}: A platform for subject-specific image analysis, visualization, and clinical support.
\newblock In \emph{Intraoperative Imaging and Image-Guided Therapy}, pages 277--289. Springer New York, 2013.

\bibitem[Kim et~al.(2019)Kim, Lee, Kim, and Lee]{kim2019effect}
Young~Jae Kim, Hyun-Ju Lee, Kwang~Gi Kim, and Seung~Hyun Lee.
\newblock The effect of {CT} scan parameters on the measurement of {CT} radiomic features: A lung nodule phantom study.
\newblock \emph{Comput. Math. Methods Med.}, 2019:\penalty0 8790694, February 2019.

\bibitem[Larue et~al.(2017)Larue, van Timmeren, de~Jong, Feliciani, Leijenaar, Schreurs, Sosef, Raat, van~der Zande, Das, van Elmpt, and Lambin]{larue2017influence}
Ruben T H~M Larue, Janna~E van Timmeren, Evelyn E~C de~Jong, Giacomo Feliciani, Ralph T~H Leijenaar, Wendy M~J Schreurs, Meindert~N Sosef, Frank H P~J Raat, Frans H~R van~der Zande, Marco Das, Wouter van Elmpt, and Philippe Lambin.
\newblock Influence of gray level discretization on radiomic feature stability for different {CT} scanners, tube currents and slice thicknesses: a comprehensive phantom study.
\newblock \emph{Acta Oncologica}, 56\penalty0 (11):\penalty0 1544--1553, 2017.

\bibitem[Li et~al.(2018)Li, Lu, Xiao, Dercle, Huang, Zhang, Schwartz, Li, and Zhao]{li2018ct}
Yajun Li, Lin Lu, Manjun Xiao, Laurent Dercle, Yue Huang, Zishu Zhang, Lawrence~H Schwartz, Daiqiang Li, and Binsheng Zhao.
\newblock {CT} slice thickness and convolution kernel affect performance of a radiomic model for predicting {EGFR} status in non-small cell lung cancer: A preliminary study.
\newblock \emph{Sci. Rep.}, 8\penalty0 (1):\penalty0 17913, December 2018.

\bibitem[Ligero et~al.(2021)Ligero, Jordi-Ollero, Bernatowicz, Garcia-Ruiz, Delgado-Muñoz, Leiva, Mast, Suarez, Sala-Llonch, Calvo, Escobar, Navarro-Martin, Villacampa, Dienstmann, and Perez-Lopez]{ligero2021minimizing}
Marta Ligero, Olivia Jordi-Ollero, Kinga Bernatowicz, Alonso Garcia-Ruiz, Eric Delgado-Muñoz, David Leiva, Richard Mast, Cristina Suarez, Roser Sala-Llonch, Nahum Calvo, Manuel Escobar, Arturo Navarro-Martin, Guillermo Villacampa, Rodrigo Dienstmann, and Raquel Perez-Lopez.
\newblock Minimizing acquisition-related radiomics variability by image resampling and batch effect correction to allow for large-scale data analysis.
\newblock \emph{Eur. Radiol.}, 31\penalty0 (3):\penalty0 1460--1470, March 2021.

\bibitem[Lin(1989)]{lin1989concordance}
Lawrence I-Kuei Lin.
\newblock A concordance correlation coefficient to evaluate reproducibility.
\newblock \emph{Biometrics}, 45\penalty0 (1):\penalty0 255--268, March 1989.

\bibitem[Lu et~al.(2016)Lu, Ehmke, Schwartz, and Zhao]{lu2016assessing}
Lin Lu, Ross~C Ehmke, Lawrence~H Schwartz, and Binsheng Zhao.
\newblock Assessing agreement between radiomic features computed for multiple {CT} imaging settings.
\newblock \emph{PLoS One}, 11\penalty0 (12):\penalty0 e0166550, December 2016.

\bibitem[Mackin et~al.(2017)Mackin, Fave, Zhang, Yang, Jones, Ng, and Court]{mackin2017harmonizing}
Dennis Mackin, Xenia Fave, Lifei Zhang, Jinzhong Yang, A~Kyle Jones, Chaan~S Ng, and Laurence Court.
\newblock Harmonizing the pixel size in retrospective computed tomography radiomics studies.
\newblock \emph{PLoS One}, 12\penalty0 (9):\penalty0 e0178524, September 2017.

\bibitem[Meyer et~al.(2019)Meyer, Ronald, Vernuccio, Nelson, Ramirez-Giraldo, Solomon, Patel, Samei, and Marin]{meyer2019reproducibility}
Mathias Meyer, James Ronald, Federica Vernuccio, Rendon~C Nelson, Juan~Carlos Ramirez-Giraldo, Justin Solomon, Bhavik~N Patel, Ehsan Samei, and Daniele Marin.
\newblock Reproducibility of {CT} radiomic features within the same patient: Influence of radiation dose and {CT} reconstruction settings.
\newblock \emph{Radiology}, 293\penalty0 (3):\penalty0 583--591, December 2019.

\bibitem[Mojtahedi et~al.(2022)Mojtahedi, Hamghalam, Do, and Simpson]{mojtahedi2022towards}
Ramtin Mojtahedi, Mohammad Hamghalam, Richard K~G Do, and Amber~L Simpson.
\newblock Towards optimal patch size in vision transformers for tumor segmentation.
\newblock In \emph{Multiscale Multimodal Medical Imaging. MMMI 2022}, pages 110--120, 2022.

\bibitem[Müllner(2011)]{mullner2011modern}
Daniel Müllner.
\newblock Modern hierarchical, agglomerative clustering algorithms.
\newblock \emph{arXiv:1109. 2378 [stat. ML]}, 2011.

\bibitem[Park et~al.(2019)Park, Lee, Do, Lee, Bae, Park, Jung, and Seo]{park2019deep}
Sohee Park, Sang~Min Lee, Kyung~Hyun Do, June~Goo Lee, Woong Bae, Hyunho Park, Kyu~Hwan Jung, and Joon~Beom Seo.
\newblock Deep learning algorithm for reducing {CT} slice thickness: Effect on reproducibility of radiomic features in lung cancer.
\newblock \emph{Korean J. Radiol.}, 20\penalty0 (10):\penalty0 1431--1440, October 2019.

\bibitem[Peoples et~al.(2023)Peoples, Hamghalam, James, Wasim, Gangai, Kang, Rong, Chun, Do, and Simpson]{peoples2023examining}
Jacob~J Peoples, Mohammad Hamghalam, Imani James, Maida Wasim, Natalie Gangai, Hyunseon~Christine Kang, Xiujiang~John Rong, Yun~Shin Chun, Richard K~G Do, and Amber~L Simpson.
\newblock Examining the effects of slice thickness on the reproducibility of {CT} radiomics for patients with colorectal liver metastases.
\newblock In \emph{Uncertainty for Safe Utilization of Machine Learning in Medical Imaging. UNSURE 2023}, pages 42--52, 2023.

\bibitem[Perrin et~al.(2018)Perrin, Midya, Yamashita, Chakraborty, Saidon, Jarnagin, Gonen, Simpson, and Do]{perrin2018short-term}
Thomas Perrin, Abhishek Midya, Rikiya Yamashita, Jayasree Chakraborty, Tome Saidon, William~R Jarnagin, Mithat Gonen, Amber~L Simpson, and Richard K~G Do.
\newblock Short-term reproducibility of radiomic features in liver parenchyma and liver malignancies on contrast-enhanced {CT} imaging.
\newblock \emph{Abdom. Radiol. (NY)}, 43\penalty0 (12):\penalty0 3271--3278, 2018.

\bibitem[Raunig et~al.(2014)Raunig, McShane, Pennello, Gatsonis, Carson, Voyvodic, Wahl, Kurland, Schwarz, Gönen, Zahlmann, Kondratovich, O'Donnell, Petrick, Cole, Garra, Sullivan, and {QIBA Technical Performance Working Group}]{raunig2014quantitative}
David~L Raunig, Lisa~M McShane, Gene Pennello, Constantine Gatsonis, Paul~L Carson, James~T Voyvodic, Richard~L Wahl, Brenda~F Kurland, Adam~J Schwarz, Mithat Gönen, Gudrun Zahlmann, Marina~V Kondratovich, Kevin O'Donnell, Nicholas Petrick, Patricia~E Cole, Brian Garra, Daniel~C Sullivan, and {QIBA Technical Performance Working Group}.
\newblock Quantitative imaging biomarkers: A review of statistical methods for technical performance assessment.
\newblock \emph{Stat. Methods Med. Res.}, 24\penalty0 (1):\penalty0 27--67, June 2014.

\bibitem[Shafiq-Ul-Hassan et~al.(2017)Shafiq-Ul-Hassan, Zhang, Latifi, Ullah, Hunt, Balagurunathan, Abdalah, Schabath, Goldgof, Mackin, Court, Gillies, and Moros]{shafiq-ul-hassan2017intrinsic}
Muhammad Shafiq-Ul-Hassan, Geoffrey~G Zhang, Kujtim Latifi, Ghanim Ullah, Dylan~C Hunt, Yoganand Balagurunathan, Mahmoud~Abrahem Abdalah, Matthew~B Schabath, Dmitry~G Goldgof, Dennis Mackin, Laurence~Edward Court, Robert~James Gillies, and Eduardo~Gerardo Moros.
\newblock Intrinsic dependencies of {CT} radiomic features on voxel size and number of gray levels.
\newblock \emph{Med. Phys.}, 44\penalty0 (3):\penalty0 1050--1062, March 2017.

\bibitem[Shafiq-ul Hassan et~al.(2018)Shafiq-ul Hassan, Latifi, Zhang, Ullah, Gillies, and Moros]{shafiq-ul-hassan2018voxel}
Muhammad Shafiq-ul Hassan, Kujtim Latifi, Geoffrey Zhang, Ghanim Ullah, Robert Gillies, and Eduardo Moros.
\newblock Voxel size and gray level normalization of {CT} radiomic features in lung cancer.
\newblock \emph{Sci. Rep.}, 8\penalty0 (1):\penalty0 10545, July 2018.

\bibitem[Simpson et~al.(2017)Simpson, Doussot, Creasy, Adams, Allen, DeMatteo, Gönen, Kemeny, Kingham, Shia, Jarnagin, Do, and D'Angelica]{simpson2017computed}
Amber~L Simpson, Alexandre Doussot, John~M Creasy, Lauryn~B Adams, Peter~J Allen, Ronald~P DeMatteo, Mithat Gönen, Nancy~E Kemeny, T~Peter Kingham, Jinru Shia, William~R Jarnagin, Richard K~G Do, and Michael~I D'Angelica.
\newblock Computed tomography image texture: A noninvasive prognostic marker of hepatic recurrence after hepatectomy for metastatic colorectal cancer.
\newblock \emph{Ann. Surg. Oncol.}, 24\penalty0 (9):\penalty0 2482--2490, 2017.

\bibitem[Simpson et~al.(2023)Simpson, Peoples, Creasy, Fichtinger, Gangai, Lasso, Keshava~Murthy, Shia, D'Angelica, and Do]{simpson2023preoperative}
Amber~L Simpson, Jacob Peoples, John~M Creasy, Gabor Fichtinger, Natalie Gangai, Andras Lasso, Krishna~Nand Keshava~Murthy, Jinru Shia, Michael~I D'Angelica, and Richard K~G Do.
\newblock Preoperative {CT} and survival data for patients undergoing resection of colorectal liver metastases ({Colorectal-Liver-Metastases) (Version 2) [Data set]}, 2023.

\bibitem[Simpson et~al.(2024)Simpson, Peoples, Creasy, Fichtinger, Gangai, Keshavamurthy, Lasso, Shia, D'Angelica, and Do]{simpson2024preoperative}
Amber~L Simpson, Jacob Peoples, John~M Creasy, Gabor Fichtinger, Natalie Gangai, Krishna~N Keshavamurthy, Andras Lasso, Jinru Shia, Michael~I D'Angelica, and Richard K~G Do.
\newblock Preoperative {CT} and survival data for patients undergoing resection of colorectal liver metastases.
\newblock \emph{Sci. Data}, 11\penalty0 (1):\penalty0 172, February 2024.

\bibitem[Traverso et~al.(2018)Traverso, Wee, Dekker, and Gillies]{traverso2018repeatability}
Alberto Traverso, Leonard Wee, Andre Dekker, and Robert Gillies.
\newblock Repeatability and reproducibility of radiomic features: A systematic review.
\newblock \emph{Int. J. Radiat. Oncol. Biol. Phys.}, 102\penalty0 (4):\penalty0 1143--1158, November 2018.

\bibitem[Vallières et~al.(2017)Vallières, Kay-Rivest, Perrin, Liem, Furstoss, Aerts, Khaouam, Nguyen-Tan, Wang, Sultanem, Seuntjens, and El~Naqa]{vallieres2017radiomics}
Martin Vallières, Emily Kay-Rivest, Léo~Jean Perrin, Xavier Liem, Christophe Furstoss, Hugo J W~L Aerts, Nader Khaouam, Phuc~Felix Nguyen-Tan, Chang-Shu Wang, Khalil Sultanem, Jan Seuntjens, and Issam El~Naqa.
\newblock Radiomics strategies for risk assessment of tumour failure in head-and-neck cancer.
\newblock \emph{Sci. Rep.}, 7\penalty0 (1):\penalty0 10117, August 2017.

\bibitem[van Griethuysen et~al.(2017)van Griethuysen, Fedorov, Parmar, Hosny, Aucoin, Narayan, Beets-Tan, Fillion-Robin, Pieper, and Aerts]{vanGriethuysen2017computational}
Joost J~M van Griethuysen, Andriy Fedorov, Chintan Parmar, Ahmed Hosny, Nicole Aucoin, Vivek Narayan, Regina G~H Beets-Tan, Jean-Christophe Fillion-Robin, Steve Pieper, and Hugo J W~L Aerts.
\newblock Computational radiomics system to decode the radiographic phenotype.
\newblock \emph{Cancer Res.}, 77\penalty0 (21):\penalty0 e104--e107, November 2017.

\bibitem[van Timmeren et~al.(2016)van Timmeren, Leijenaar, van Elmpt, Wang, Zhang, Dekker, and Lambin]{vanTimmeren2016test-retest}
Janna~E van Timmeren, Ralph T~H Leijenaar, Wouter van Elmpt, Jiazhou Wang, Zhen Zhang, André Dekker, and Philippe Lambin.
\newblock Test-retest data for radiomics feature stability analysis: Generalizable or study-specific?
\newblock \emph{Tomography}, 2\penalty0 (4):\penalty0 361--365, December 2016.

\bibitem[Varghese et~al.(2019)Varghese, Hwang, Cen, Levy, Liu, Lau, Rivas, Desai, Goodenough, and Duddalwar]{varghese2019reliability}
Bino~A Varghese, Darryl Hwang, Steven~Y Cen, Joshua Levy, Derek Liu, Christopher Lau, Marielena Rivas, Bhushan Desai, David~J Goodenough, and Vinay~A Duddalwar.
\newblock Reliability of {CT}-based texture features: Phantom study.
\newblock \emph{J. Appl. Clin. Med. Phys.}, 20\penalty0 (8):\penalty0 155--163, August 2019.

\bibitem[Ward(1963)]{ward1963hierarchical}
Joe~H Ward.
\newblock Hierarchical grouping to optimize an objective function.
\newblock \emph{J. Am. Stat. Assoc.}, 58\penalty0 (301):\penalty0 236--244, March 1963.

\bibitem[Wilcoxon(1945)]{wilcoxon1945individual}
Frank Wilcoxon.
\newblock Individual comparisons by ranking methods.
\newblock \emph{Biom. Bull.}, 1\penalty0 (6):\penalty0 80, December 1945.

\bibitem[Xu et~al.(2022)Xu, Lu, Sun, E, Lian, Yang, Schwartz, Yang, and Zhao]{xu2022effect}
Yan Xu, Lin Lu, Shawn~H Sun, Lin-Ning E, Wei Lian, Hao Yang, Lawrence~H Schwartz, Zheng-Han Yang, and Binsheng Zhao.
\newblock Effect of {CT} image acquisition parameters on diagnostic performance of radiomics in predicting malignancy of pulmonary nodules of different sizes.
\newblock \emph{Eur. Radiol.}, 32\penalty0 (3):\penalty0 1517--1527, March 2022.

\bibitem[Yang et~al.(2021)Yang, Wu, Zhang, and Li]{yang2021evaluation}
Shouxin Yang, Ning Wu, Li~Zhang, and Meng Li.
\newblock Evaluation of the linear interpolation method in correcting the influence of slice thicknesses on radiomic feature values in solid pulmonary nodules: a prospective patient study.
\newblock \emph{Ann. Transl. Med.}, 9\penalty0 (4):\penalty0 279, February 2021.

\bibitem[Zhao(2021)]{zhao2021understanding}
Binsheng Zhao.
\newblock Understanding sources of variation to improve the reproducibility of radiomics.
\newblock \emph{Front. Oncol.}, 11, 2021.

\bibitem[Zhao et~al.(2014)Zhao, Tan, Tsai, Schwartz, and Lu]{zhao2014exploring}
Binsheng Zhao, Yongqiang Tan, Wei~Yann Tsai, Lawrence~H Schwartz, and Lin Lu.
\newblock Exploring variability in {CT} characterization of tumors: A preliminary phantom study.
\newblock \emph{Transl. Oncol.}, 7\penalty0 (1):\penalty0 88--93, February 2014.

\bibitem[Zhao et~al.(2016)Zhao, Tan, Tsai, Qi, Xie, Lu, and Schwartz]{zhao2016reproducibility}
Binsheng Zhao, Yongqiang Tan, Wei-Yann Tsai, Jing Qi, Chuanmiao Xie, Lin Lu, and Lawrence~H Schwartz.
\newblock Reproducibility of radiomics for deciphering tumor phenotype with imaging.
\newblock \emph{Scientific Reports}, 6\penalty0 (1), 2016.

\bibitem[Zwanenburg et~al.(2016)Zwanenburg, Leger, Vallières, and Löck]{zwanenburg2016image}
Alex Zwanenburg, Stefan Leger, Martin Vallières, and Steffen Löck.
\newblock Image biomarker standardisation initiative: Reference manual.
\newblock \emph{arXiv:1612. 07003 [cs. CV]}, 2016.

\bibitem[Zwanenburg et~al.(2020)Zwanenburg, Vallières, Abdalah, Aerts, Andrearczyk, Apte, Ashrafinia, Bakas, Beukinga, Boellaard, Bogowicz, Boldrini, Buvat, Cook, Davatzikos, Depeursinge, Desseroit, Dinapoli, Dinh, Echegaray, Naqa, Fedorov, Gatta, Gillies, Goh, Götz, Guckenberger, Ha, Hatt, Isensee, Lambin, Leger, Leijenaar, Lenkowicz, Lippert, Losnegård, Maier-Hein, Morin, Müller, Napel, Nioche, Orlhac, Pati, Pfaehler, Rahmim, Rao, Scherer, Siddique, Sijtsema, Fernandez, Spezi, Steenbakkers, Tanadini-Lang, Thorwarth, Troost, Upadhaya, Valentini, van Dijk, van Griethuysen, van Velden, Whybra, Richter, and Löck]{zwanenburg2020image}
Alex Zwanenburg, Martin Vallières, Mahmoud~A Abdalah, Hugo J W~L Aerts, Vincent Andrearczyk, Aditya Apte, Saeed Ashrafinia, Spyridon Bakas, Roelof~J Beukinga, Ronald Boellaard, Marta Bogowicz, Luca Boldrini, Irène Buvat, Gary J~R Cook, Christos Davatzikos, Adrien Depeursinge, Marie-Charlotte Desseroit, Nicola Dinapoli, Cuong~Viet Dinh, Sebastian Echegaray, Issam~El Naqa, Andriy~Y Fedorov, Roberto Gatta, Robert~J Gillies, Vicky Goh, Michael Götz, Matthias Guckenberger, Sung~Min Ha, Mathieu Hatt, Fabian Isensee, Philippe Lambin, Stefan Leger, Ralph T~H Leijenaar, Jacopo Lenkowicz, Fiona Lippert, Are Losnegård, Klaus~H Maier-Hein, Olivier Morin, Henning Müller, Sandy Napel, Christophe Nioche, Fanny Orlhac, Sarthak Pati, Elisabeth A~G Pfaehler, Arman Rahmim, Arvind U~K Rao, Jonas Scherer, Muhammad~Musib Siddique, Nanna~M Sijtsema, Jairo~Socarras Fernandez, Emiliano Spezi, Roel J H~M Steenbakkers, Stephanie Tanadini-Lang, Daniela Thorwarth, Esther G~C Troost, Taman Upadhaya, Vincenzo Valentini, Lisanne~V van
  Dijk, Joost van Griethuysen, Floris H~P van Velden, Philip Whybra, Christian Richter, and Steffen Löck.
\newblock The image biomarker standardization initiative: Standardized quantitative radiomics for high-throughput image-based phenotyping.
\newblock \emph{Radiology}, 295\penalty0 (2):\penalty0 328--338, 2020.

\end{thebibliography}
